\documentclass[journal]{IEEEtran}

\usepackage{mathrsfs}
\usepackage[noadjust]{cite}
\usepackage{graphicx,color,overpic,psfrag}
\usepackage{amsmath, amssymb}
\usepackage{latexsym}
\usepackage{bm}
\usepackage{amssymb}
\usepackage{cases}
\usepackage{array}
\usepackage{fancyhdr}
\usepackage{setspace}

\ifCLASSOPTIONcompsoc
\usepackage[caption=false,font=normalsize,labelfont=sf,textfont=sf]{subfig}
\else
\usepackage[caption=false,font=footnotesize]{subfig}
\fi

\usepackage{url}
\usepackage{algpseudocode}
\usepackage{algorithm}
\usepackage{blkarray}
\usepackage{booktabs}

\usepackage{multirow}
\usepackage{dsfont}
\usepackage{tabularx}
\usepackage[table]{xcolor}

\usepackage{amsfonts}

\usepackage{letltxmacro}

\graphicspath{{figure/}}

\newcommand{\ra}[1]{\renewcommand{\arraystretch}{#1}}

\newtheorem{prop}{Proposition}

\newcommand{\figref}[1]{Fig. \ref{#1}}
\newcommand{\tabref}[1]{Table \ref{#1}}
\newcommand{\alref}[1]{Algorithm \ref{#1}}
\newcommand{\appref}[1]{Appendix \ref{#1}}
\newcommand{\secref}[1]{Section \ref{#1}}

\newcommand{\propref}[1]{{Proposition \ref{#1}}}
\newcommand{\subfigref}[2]{Fig. \ref{#1}\subref{#2}}
\newcommand{\subsecref}[1]{Section \ref{#1}}

\newcommand{\tr}[1]{\mathsf{tr}\left\{#1\right\}}

\newcommand{\argmax}[1]{\mathop{\arg\max}\limits_{#1}}
\newcommand{\argmin}[1]{\mathop{\arg\min}\limits_{#1}}

\newcommand{\cP}{\mathcal{P}}

\newcommand{\bb}{\mathbf{b}}
\newcommand{\bc}{\mathbf{c}}

\newcommand{\bn}{\mathbf{n}}

\newcommand{\bt}{\mathbf{t}}
\newcommand{\bu}{\mathbf{u}}
\newcommand{\bv}{\mathbf{v}}

\newcommand{\bx}{\mathbf{x}}
\newcommand{\by}{\mathbf{y}}
\newcommand{\bz}{\mathbf{z}}

\newcommand{\bA}{\mathbf{A}}
\newcommand{\bB}{\mathbf{B}}
\newcommand{\bC}{\mathbf{C}}

\newcommand{\bE}{\mathbf{E}}
\newcommand{\bF}{\mathbf{F}}
\newcommand{\bG}{\mathbf{G}}
\newcommand{\bH}{\mathbf{H}}
\newcommand{\bI}{\mathbf{I}}

\newcommand{\bP}{\mathbf{P}}
\newcommand{\bQ}{\mathbf{Q}}
\newcommand{\bR}{\mathbf{R}}
\newcommand{\bS}{\mathbf{S}}
\newcommand{\bT}{\mathbf{T}}
\newcommand{\bU}{\mathbf{U}}
\newcommand{\bV}{\mathbf{V}}
\newcommand{\bW}{\mathbf{W}}

\newcommand{\emath}{\mathrm{e}}

\newcommand{\bSigma}{{\boldsymbol\Sigma}}

\newcommand{\bLambda}{{\boldsymbol\Lambda}}

\newcommand{\bXi}{{\boldsymbol\Xi}}

\newcommand{\ntb}{\notag\\}

\newcolumntype{L}{>{\hspace*{-\tabcolsep}}l}
\newcolumntype{R}{c<{\hspace*{-\tabcolsep}}}
\definecolor{lightblue}{rgb}{0.93,0.95,1.0}

\newcommand{\R}{\mathbb{R}}
\newcommand{\C}{\mathbb{C}}
\newcommand{\Q}{\mathbf{Q}}

\newcommand{\I}{\mathbf{I}}

\newcommand{\K}{\cal{K}}

\newcommand{\Gk}{\mathbf{G}_{k}}

\newcommand{\Kkopt}{\mathbf{K}_{k,\mathrm{opt}}}
\newcommand{\Uk}{\mathbf{U}_{k}}

\newcommand{\Qk}{\mathbf{Q}_k}

\newcommand{\Qkopt}{\mathbf{Q}_{k,\mathrm{opt}}}

\newcommand{\Rki}{\mathbf{R}_{k,i}}

\newcommand{\lambdaki}{\lambda_{k,i}}

\newcommand{\Pmaxk}{P_{\mathrm{max},k}}
\newcommand{\Pmax}{P_{\mathrm{max}}}

\newcommand{\bps}{\mathrm {bps}}

\newcommand{\Hz}{\mathrm {Hz}}

\newcommand{\etaSE}{{\eta}_{\mathrm{SE}}}

\newcommand{\etaSEb}{{\overline {\eta}}_{\mathrm{SE}}}

\newcommand{\Lambdak}{\bm{\Lambda}_{k}}
\newcommand{\Gammak}{\bm{\Gamma}_{k}}
\newcommand{\muk}{\mu_k}

\newcommand{\lambdakiopt}{\lambda_{k,i,\mathrm{opt}}}
\newcommand{\mukopt}{\mu_{k,\mathrm{opt}}}

\newcommand{\sumkeqtoK}{\sum\limits_{k=1}^{K}}

\newcommand{\sigmatwo}{\sigma^{2}}
\newcommand{\Psik}{\mathbf{\Psi}_{k}}
\newcommand{\psik}{\bm{\psi}_{k}}
\newcommand{\gammak}{\bm{\gamma}_{k}}

\newcommand{\bPsi}{\mathbf{\Psi}}

\newcommand{\bGamma}{\bm{\Gamma}}

\newcommand{\Htwok}{\bH_{2,k}}
\newcommand{\bPhi}{\bm{\Phi}}
\newcommand{\bphi}{\bm{\phi}}

\newcommand{\Vtildeone}{\widetilde{\bV}_1}
\newcommand{\Xitilde}{\widetilde{\bXi}}
\newcommand{\sumieqtoAk}{\sum\limits_{i=1}^{A_k}}
\newcommand{\lambdamax}{\lambda_{\mathrm{max}}}
\newcommand{\bDelta}{\bm{\Delta}}
\newcommand{\Htwoktilde}{\widetilde{\bH}_{2,k}}
\newcommand{\Omegatwok}{\bm{\Omega}_{2,k}}
\allowdisplaybreaks

\begin{document}


\title{Hybrid RIS and DMA Assisted Multiuser MIMO Uplink Transmission With Electromagnetic Exposure Constraints}

\author{
Hanyu~Jiang, Li~You, Jue~Wang, Wenjin~Wang, and Xiqi~Gao%

\thanks{
	Copyright (c) 2015 IEEE. Personal use of this material is permitted. However, permission to use this material for any other purposes must be obtained from the IEEE by sending a request to pubs-permissions@ieee.org.
}
\thanks{
	Part of this work was submitted to Globecom 2022 SPC \cite{Globcom2022}.
}
\thanks{
	Hanyu Jiang, Li You, Wenjin Wang, and Xiqi Gao are with the National Mobile Communications Research Laboratory, Southeast University, Nanjing 210096, China (e-mail: hyjiang@seu.edu.cn; lyou@seu.edu.cn; wangwj@seu.edu.cn; xqgao@seu.edu.cn).
	
	Jue Wang is with School of Information Science and Technology, Nantong University, Nantong 226019, China (e-mail: wangjue@ntu.edu.cn). 
		
}
}

\maketitle

\begin{abstract}
In the fifth-generation and beyond era, reconfigurable intelligent surface (RIS) and dynamic metasurface antennas (DMAs) are emerging metamaterials keeping up with the demand for high-quality wireless communication services, which promote the diversification of portable wireless terminals. However, along with the rapid expansion of wireless devices, the electromagnetic (EM) radiation increases unceasingly and inevitably affects public health, which requires a limited exposure level in the transmission design. To reduce the EM radiation and preserve the quality of communication service, we investigate the spectral efficiency (SE) maximization with EM constraints for uplink transmission in hybrid RIS and DMA assisted multiuser multiple-input multiple-output systems. Specifically, alternating optimization is adopted to optimize the transmit covariance, RIS phase shift, and DMA weight matrices. We first figure out the water-filling solutions of transmit covariance matrices with given RIS and DMA parameters. Then, the RIS phase shift matrix is optimized via the weighted minimum mean square error, block coordinate descent and minorization-maximization methods. Furthermore, we solve the unconstrainted DMA weight matrix optimization problem in closed form and then design the DMA weight matrix to approach this performance under DMA constraints. Numerical results confirm the effectiveness of the EM aware SE maximization transmission scheme over the conventional baselines.

\end{abstract}

%

\section{Introduction}
Along with the increasing demand for the quality of communication service, future wireless systems are required to support a peak rate of thousands of megabits per second and service density of hundreds of multi-antenna devices per square meter \cite{ZBM20}. To this end, a large number of wireless terminals and base stations (BSs) will be deployed for greater data traffic \cite{WSN17}. However, these ubiquitous communication services are often accompanied by great challenges, in not only technical implementations but also environmental considerations. On the one hand, the deployment of plenty of BSs with a large number of radio frequency (RF) chains will immensely increase the power consumption and impose a great burden on the shape design of antenna arrays \cite{ZWX17,ZZW19}. On the other hand, the surge in the number of network connections engenders the substantial growth of electromagnetic (EM) radiation that is nonnegligible to the public health \cite{JHB20}. 

To reduce the hardware cost of deploying large-scale antennas at the BS, we resort to the dynamic metasurface antenna (DMA). DMA is proposed as a brand-new concept for the realization of antenna arrays, where metamaterials are used in the aperture antenna design \cite{SAI21,YXA21}. A simple DMA-based array is composed of several microstrips in parallel, each of which consists of a set of subwavelength and frequency selective resonant metamaterial elements \cite{HDM13}. They are capable of tailoring the beams and processing signals in a dynamically configurable way \cite{SIX16}. The application of DMAs enables a large number of adjustable elements, which are reconfigurable owing to the introduction of solid-state switchable components in each metamaterial, to be set in a small physical area \cite{SDE19}. In addition, the number of RF chains required for DMA assisted communication is equal to that of microstrips, which is usually far less than that required in conventional antenna systems \cite{YXA21}. Hence, both the physical size and the power consumption will be greatly reduced.
Instead of a passive configurable metasurface that only reflects the signals, DMA array performs as an active transceiver that inherently implements signal processing techniques such as analog beamforming and combining \cite{WSE21}.
The flexible architecture of metasurface as an active antenna array makes DMAs attractive for multiple-input multiple-output (MIMO) transceivers in future wireless networks \cite{SAI21}. 

However, the application of DMAs usually restricts the achievable system spectral efficiency (SE) due to the reduction of the RF chains at the BS. To compensate for this defect, adopting reconfigurable intelligent surface (RIS) is proposed as an effective method to improve the system SE with low hardware complexity \cite{ZLP21}. 
Unlike DMAs which perform as active metamaterials equipped at BS, RIS is a two-dimensional metamaterial surface composed of ultra-thin composite material layers that can programmatically reflect the incident EM waves to the desired directions \cite{LNT18,AZW20}.
RIS contains a plurality of reflecting elements that are usually constituted by positive-intrinsic-negative diodes to tune the phase of the incident signal in a software defined manner \cite{CAZ21}. With the reconfigurable intelligent property, RIS can superimpose the incident waves by adjusting the phase shifts and then reflect them to the appropriate directions, which provides excellent flexibility for cellular mobile communications with complex propagating scenes \cite{MA21,CMH21,ZDS21}. Consequently, by optimizing the phase shifts of RIS reflecting elements in the system design, it is possible to enhance the designed signal power while suppressing interference so as to improve the system SE. In addition, with low hardware footprints, RIS assisted communication has become a valuable wireless transmission strategy in the next generation network \cite{ZLP21,CAZ21}. 

As for the concern that the public is vulnerable to the increasing EM radiation, EM exposure is quantified at the user side and specified at a low level by communication regulatory agencies, which calls for new transmission strategy designs for MIMO uplink \cite{CLL20,C04}. 
EM exposure refers to the radiation exposure generated by the propagation of EM waves, which usually comes from the power electronic devices and various kinds of artificial and natural lights \cite{JHB20}. Recently, the swift development of wireless networks and the gradual maturity of the Internet of Things technology have made EM exposure a critical issue \cite{AGM15}. Therefore, many government departments require that the EM radiation emitted by qualified electronics be kept at a low dose. To this end, specific absorption rate (SAR), which denotes the absorbed power per unit mass of human tissues with the unit W/kg, has become a standard metric to measure the exposure level of the public \cite{C04}. According to the Federal Communications Commission (FCC) standard, for wireless devices with frequencies in the range of 100 kHz to 6 GHz, the peak SAR on partial-body EM exposure should be limited to 1.6 W/kg \cite{SARstandard}. 
As a time-averaged quadratic metric, SAR mainly relates to the near-field of transmitting antenna in uplink communication, where the peak value of SAR is much higher than the average \cite{HLY13}. Then, SAR is considered to comply with the worst-case. In single antenna cases, SAR can be naturally complied for the worst-case by reducing the transmit power. However, for multi-antenna systems, it is inefficient to ensure the worst-case compliance in the same way, which brings the demand for the transmitter adaptive design that is actively satisfying different SAR constraints \cite{YLH15,JYW22}.

In the fifth-generation (5G) cellular systems, the prevalence of wireless handsets with multiple antennas has evoked the EM aware optimization design for the transmission SE. For example, authors in \cite{HLY14} proposed the matrix constraint form of EM radiation in the uplink transmission design with multi-antenna user terminals. Then, the SAR constrained sum-rate maximization precoding at users was investigated in \cite{YLH17} for the uplink multiuser MIMO systems. Recently, due to the proposal of the controllable intelligent radio environment \cite{HHA20}, RIS and DMA have been paid enormous attention to the next generation wireless networks. In \cite{YXH21,YXN21}, the phase shifts of RIS and transmit precoding at BS were jointly optimized to obtain the maximum energy efficiency (EE) and resource efficiency of the downlink multiuser MIMO system. In addition to the RIS assisted system, which can dynamically adjust the propagation environment, DMAs can be adopted at the BS to reduce the energy consumption and implementation cost of massive antenna arrays. The impact of DMAs on the capacity of wireless systems was investigated in \cite{SDE19,YXA21}, where the corresponding weights of DMAs at the BS are optimized to maximize SE and EE, respectively.
Note that although most studies focus on the power allocation algorithms of RIS or DMA assisted systems under power constraints, the introduction of EM exposure constraints poses new challenges to the optimization. Furthermore, the fast time-varying channel is a common scenario in wireless communication systems, where instantaneous channel state information (CSI) is difficult to obtain and becomes outdated easily \cite{GJL09,WMJ13}. Compared with instantaneous CSI, the statistical CSI, e.g., the spatial correlation and channel mean, is more likely to be stable for a longer period, thus bringing the lower cost for acquisition. In this case, efficient utilization of statistical CSI is promising for transmission design.
In addition, the CSI is usually not perfectly available in practical systems, which might degrade the transmission performance. Robust transmission design which incorporates the imperfect CSI effect is of practical interest \cite{NCN21}.

In this paper, we investigate the EM aware SE maximization design in RIS and DMA assisted multiuser MIMO uplink transmission. Specifically, the transmit precoding, RIS phases, and DMA weights are jointly designed to maximize the system SE under both power and SAR constraints at users. We consider the practical scenario where RIS and DMAs are statically deployed, and hence full CSI between RIS and DMA is available in the considered system. On the other hand, both full CSI and partial CSI cases from users to RIS are considered in the optimization design. We intend to figure out the impact of EM exposure on the SE performance of hybrid RIS and DMA assisted systems by comparing with conventional systems, then compare our proposed algorithm with the conventional backoff algorithms. The main contributions of this paper are outlined as follows:

\begin{itemize}	
	\item[$\bullet$] We investigate the hybrid RIS and DMA assisted multiuser MIMO system for practical interest, where RIS and DMA are actually complementary in wireless transmissions. In particular, RIS is adopted to dynamically adjust the propagation environment. Meanwhile, DMA is adopted as a new form of BS antennas to reduce the energy consumption and implementation cost. 
	\item[$\bullet$] 
	The SAR constraints are taken into account to protect users from the high dose of EM radiation in hybrid RIS and DMA assisted transmissions. To address the EM aware problem, we propose a modified water-filling algorithm to optimize the transmit covariances, apply the minimum mean square error (MMSE), block coordinate descent (BCD) and minorization-maximization (MM) methods to optimize RIS phase shifts in closed form, and design the DMA weights by approaching the performance of unconstrainted DMA problems.
	\item[$\bullet$] The partial CSI case is studied in the transmission scenario. To reduce the complexity of the Monte Carlo method in dealing with partial CSI, we apply the deterministic equivalent (DE) method to asymptotically approximate SE. Then, we propose the AO-based SE maximization algorithm with the utilization of the channel eigenmode coupling matrices from users to RIS.
\end{itemize}

The rest of the paper is organized as follows: \secref{sec:system model} elaborates the model of RIS and DMA assisted system, specifies the representation of EM exposure, and formulates the problem of SE maximization for two cases of available CSI. Based on the AO framework, \secref{sec:opt_SE_design_PCSI} provides SE maximization algorithms of the optimization variables separately under full CSI scenario. \secref{sec:opt_SE_design_SCSI} address the same problems with partial CSI. \secref{sec:analysis} analyzes the convergence and complexity of the overall AO-base algorithm. In \secref{sec:numerical_results}, numerical results are presented to analyze the performance of the proposed algorithms. Lastly, \secref{sec:conclusion} concludes the study.

\emph{Notations}: Suppose matrix $\bA = \mathrm{diag}\{\bA_k\}_{k=1}^{K}$ is the block diagonal matrix composed of $K$ sub-matrices, and the element on the $k$th diagonal block sequenced from the upper left is $\bA_k$. $\bA_{[1:k]}$ is the matrix obtained by truncating the first $k$ column vectors of matrix $\bA$. $\bA_{m,n}$ denotes the element located in row $m$ and column $n$ of matrix $\bA$. $\mathbb{E}\{\cdot\}$ means calculating the expected value. $\mathbf{0}$ denotes zero vector. $\C$, $\R$, $\R^+$ represent complex, real and positive real number sets,  respectively. $\odot$ denotes the Hadamard product. $\tr{\cdot}$ means the trace. $(x)^+=\max\{x,0\}$. $\Re\{\cdot\}$ means taking the real part of a complex number. $||\cdot||_\mathrm{F}^2$ is the Frobenius norm. $\jmath=\sqrt{-1}$ is the imaginary unit.

\section{System Model}\label{sec:system model}
\begin{figure}[t]
	\centering
	\includegraphics[width=0.5\textwidth]{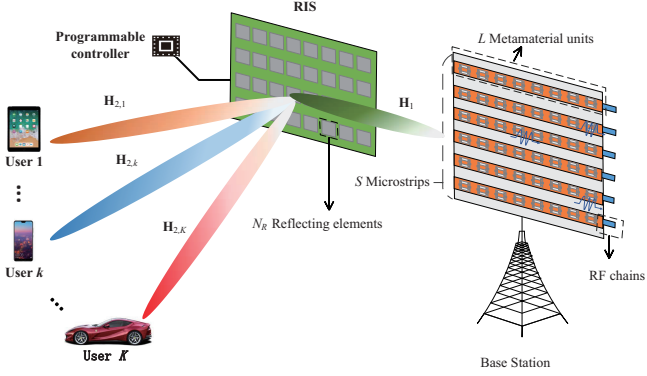}
	\caption{The hybrid RIS and DMA assisted multiuser MIMO system.}
	\label{fig:RISDMAmodel}
\end{figure}

Consider the hybrid RIS and DMA assisted multiuser MIMO uplink transmission with $K$ users in the single cell transmitting signals to a $M$-antenna BS simultaneously, as shown in \figref{fig:RISDMAmodel}. Denote the user set as $\mathcal{K}=\{1,...,K\}$, and the number of antennas for user $k\in\mathcal{K}$ is $N_k$. We assume that the encoded transmit signal from user $k$ is $\bx_k\in\C^{N_K\times 1}$, which is zero mean and independent of signals from other users, i.e., $\mathbb{E}\{\bx_k\}=\mathbf{0}, \forall k\in\mathcal{K}$ and $\mathbb{E}\{\bx_i\bx_j^H\}=\mathbf{0},\forall i \neq j \in \K$. Denote the covariance matrix corresponding transmit signal $\bx_k$ as $\Qk\triangleq\mathbb{E}\{\bx_k\bx_k^H\},\forall k \in \K$. As the elements of $\bx_k$ are spatially correlated, $\bQ_k$ is essentially a non-diagnoal matrix. 

\subsection{RIS Assised Model}
In \figref{fig:RISDMAmodel}, the signals from transmitters are reflected in the channel with the deployment of RIS consisting of $N_R$ reflecting elements, each of which can tune the phase of the incident signal separately by applying the programmable controller \cite{CMH21, AZW20}. 
As the enormous increase of the RF weakens the diffraction and scattering effect, electromagnetic waves are prone to blockage by obstacles such as buildings in urban areas \cite{HZA19}. In this paper, we assume the typical model that the direct channel from users to the BS is blocked, i.e., only the paths that users to the BS via RIS are considered in our system \cite{HZA19}.
In addition, the cases that signals experience multiple reflections by RIS are ignored due to the tremendous path loss \cite{PRW20,AZW20}. We suppose the channel matrix from user $k$ to RIS as $\Htwok\in\C^{N_R\times N_k}$, and that from the RIS to BS as $\bH_1\in\C^{M\times N_R}$. Then, the received signal gathered at the BS side can be formulized as
\begin{align}\label{equ:received_y1}
\by=\sumkeqtoK\bH_1\bPhi\Htwok\bx_k+\bn \in \C^{M\times1},
\end{align}
where $\bn$ is the additive noise following $\mathcal{CN}(0,\sigmatwo\bI_{M})$, and $\bPhi=\mathrm{diag}\{\phi_1,...,\phi_{N_R}\}$ denotes the phase shift matrix of RIS, whose diagonal elements are the reflection coefficients.

Suppose that RIS can achieve total reflection, which means for any $n\in\{1,...,N_R\}$, the reflection coefficients can be written as $\phi_n=\emath^{\jmath\theta_n}$, where $\theta_n$ is the phase shift introduced by the $n$th element of RIS. In addition, we adopt the ideal assumption that the reflecting elements can perform continuous phase shift, i.e., \cite{WLW20}
\begin{align}\label{equ:phase shift}
\phi_n\in\mathcal{F}_1\triangleq\{\phi|\phi=\emath^{\jmath\theta},\theta\in\left[\right.0,2\pi\left.\right)\},
\end{align}
where $\mathcal{F}_1$ denotes the feasible set of the reflection coefficients.

\subsection{DMA Assised Model}
Suppose that the DMA array is equipped at the BS consisting of $M$ metamaterial units. These DMAs are composed of $S$ microstrips, each of which contains $L$ metamaterial units, i.e., $M = S \cdot L$. In practice, DMA array can be regarded as a two-dimensional antenna array composed of a set of one-dimensional microstrips, and its configurable weight matrix $\bXi\in\C^{S\times M}$ can be written as \cite{SDE19}
\begin{align}\label{equ:Xi_structure}
\bXi_{s_1,(s_2-1)L+l}=
\left\{
\begin{array}{cccc}
\xi_{s_1,l}\in \mathcal{F}_2  &s_1=s_2 \\
0 &s_1\neq s_2 \\
\end{array}
\right. \in \mathcal{F}_3^{S\times M},
\end{align}
where $s_1,s_2\in\{1,...,S\}$, $l\in\{1,...,L\}$, $\{\xi_{s_1,l}\}_{\forall s_1,l}$ are the weights of the DMA elements and the feasible set of weight matrices is denoted as $\mathcal{F}_3^{S\times M}$. Please note that $\xi_{s_1,l}$ often satisfies certain constraints, e.g., its feasible set represented by $\mathcal{F}_2$ may have the following forms \cite{SYM17}:
\begin{itemize}
	\item[(1).] $\mathcal{F}_2= \C$ for unconstrained DMA weights.
	\item[(2).] $\mathcal{F}_2=[x,y]$ where $0<x<y\in\R$ for amplitude only.
	\item[(3).] $\mathcal{F}_2=c\cdot\{0,1\}$ where $c\in\R^+$ for binary amplitude.
	\item[(4).] $\mathcal{F}_2=\left\{\frac{\jmath+\emath^{\jmath\varphi}}{2}|\varphi\in\left[\right.0,2\pi\left.\right)\right\}$	for Lorentzian phase.
\end{itemize}
As shown in \figref{fig:RISDMAmodel}, the input of DMAs is comprised of the received signal $\by\in\C^{S\cdot L \times 1}$ in \eqref{equ:received_y1} at the BS side. Consider the case where the response of metamaterial elements is frequency flat as adopted in \cite{ZSG21}, the process that the signal propagates inside the corresponding microstrips can be modeled as an equivalent causal filter with the finite impulse response, and the corresponding taps are expressed by \cite{ZSG21}
\begin{align}\label{equ:response_f}
f_{s,l}=\emath^{-r_{s,l}(\alpha_s+\jmath\beta_s)},\ \forall s\in\{1,...,S\},\ l\in\{1,...,L\},
\end{align}
where $r_{s,l}$ denotes the location of the $l$th meta-material unit in the $s$th microstrip, $\alpha_s$ and $\beta_s$ represent the waveguide attenuation coefficient and the wavenumber, respectively. Note that \eqref{equ:response_f} is based on the assumption that the elements of each microstrip do not perturb the waveguide mode and do not interact with each other \cite{SYM17}. Then, the mutual coupling effect inside the microstrip is weak and can be ignored for simplicity \cite{SYM17}. 
Denote $\bF\in\C^{M\times M}$ as the designed matrix form of the equivalent impulse response where $(\bF)_{(s-1)L+l,(s-1)L+l}=f_{s,l}$. Then, the output radiation observed on DMAs can be expressed as \cite{ZSG21}
\begin{align}\label{equ:received_z}
\bz=\bXi\bF\by \in \C^{S\times 1}.
\end{align} 
It is worth emphasizing that the output of each microstrip is obtained as the linear combination of the radiation observed by the metamaterial elements of the corresponding microstrip.
In this paper, we consider the scenario that all the metamaterial elements introduce the same frequency response, i.e., $f_{s,l}=f,\ \forall s,l$ and $\bF={\bI_{M}\cdot f}$, as that in \cite{SDE19}. By combining \eqref{equ:received_y1} into \eqref{equ:received_z}, the input-output relationship of the system between the transmit signals and DMAs output can be written as
\begin{align}\label{equ:relationship_zx}
\bz=f\sumkeqtoK\bXi\bH_1\bPhi\Htwok\bx_k+\tilde{\bn} \in \C^{S\times 1},
\end{align} 
where $\tilde{\bn}=f\bXi\bn$ is the equivalent channel noise.

\subsection{System SE}
The maximum achievable SE of the communication is related to the CSI available in the considered system. In the following, we consider two scenarios as follows:
\begin{itemize}
	\item[\emph{(a).}] \emph{Full CSI}: $\bH_1$, $\{\Htwok\}_{\forall k}$ are available as the perfect instantaneous CSI in transmission.
	\item[\emph{(b).}] \emph{Partial CSI}: The considered system has access to the instantaneous CSI for the RIS-to-DMAs channel $\bH_1$, but only the statistical CSI for the users-to-RIS channels $\{\Htwok\}_{\forall k}$.	
\end{itemize}

Note that the second scenario is realistic in situations that RIS-to-DMAs channel is slowly time-varying, whereas the users-to-RIS channels are rapidly time-varying \cite{YXN21}. According to \cite{SDE19}, the general SE model encompassing these scenarios can be expressed as
\begin{align}\label{equ:SE_model}
&\etaSE(\Q, \bPhi, \bXi)=\mathbb{E}_{\{\Htwok\}}\Bigg\{\log\det\Bigg(\bI_{S}+\frac{1}{\sigmatwo}\sumkeqtoK\bXi\cdot\ntb
&\bH_1\bPhi\Htwok\Qk\Htwok^H\bPhi^H\bH_1^H\bXi^H(\bXi\bXi^H)^{-1}\Bigg)\Bigg\}\ \bps/\Hz,
\end{align}
where $\bQ\triangleq\mathrm{diag}\{\bQ_k\}_{k=1}^{K}$ represents the aggregated covariance matrices. When perfect knowledge of CSI for all channels is available, $\{\Htwok\}_{\forall k}$ in \eqref{equ:SE_model} become determined values and the expectation therein can be removed. On the other hand, when we consider the system SE with partial CSI, such as scenario (b), \eqref{equ:SE_model} degenerates into the representation of ergodic achievable SE with statistical CSI of $\{\Htwok\}_{\forall k}$.

\subsection{Electromagnetic Exposure Model}
In practical wireless uplink communications, both the power and EM exposure level can place restrictions on the transmission rate from users to the BS \cite{YLH17}. Generally, the constraints exerted on the power consumption are expressed as $\tr{\Qk}\leq \Pmaxk,\ \forall k\in\mathcal{K}$, where $\Pmaxk$ denotes the power budget of the $k$th user. In addition, the EM exposure at the users is usually measured by the SAR, which can be modeled as a quadratic function of the transmitted signal $\bx_k$ \cite{HLY13}. Typically, due to the variation of mass density and conductivity for different parts of tissues, there will be multiple SAR constraints for a single user \cite{YLH15}. For the transmitter with multiple antennas, we can model the SAR with the time-averaged quadratic constraints given by \cite{YLH15}
\begin{align}\label{equ:SAR_ki}
\mathrm{SAR}_{k,i}&=\mathbb{E}\{\tr{\bx^H_k\Rki\bx_k}\} \ntb
&=\tr{\Rki\Qk}\leq D_{k,i},\quad i\in \{1,2,...,A_k\},
\end{align}
where $A_k$ denotes the number of SAR constraints for user $k$, $D_{k,i}$ is the maximum SAR value that the $i$th body part of the $k$th user can be exposed to EM field, and $\Rki$ is the $i$th SAR matrix of user $k$ describing the radiation coefficients of transmit signals with the unit of each entry as $\mathrm{kg}^{-1}$. Commonly, the SAR matrix is dependent on the conductivity and frequency, as well as the employed instrument that dictates the surrounding boundary conditions \cite{LCA13}. However, work in \cite{CCM04} shows that the SAR measurements do not vary significantly over a fairly wide range of carrier frequencies (1.8 -- 2 GHz). Note that for narrowband communication with a bandwidth less than 200 MHz, SAR measurements are almost the same for the given angle of antennas so that only a single pointwise SAR matrix is required for an orthogonal frequency division multiplexing system. Therefore, it is reasonable to assume all the sub-carriers in the systems share the same SAR matrix. 

\subsection{Problem Formulation}
In this paper, we investigate strategy design for the RIS and DMA assisted multiuser MIMO uplink transmission with EM exposure constraints, where we aim to optimize the adjustable parameters $\{\Qk\}_{\forall k}$, $\bPhi$ and $\bXi$ to maximize the system SE. Applying the relevant constraints to \eqref{equ:SE_model}, our work can be formulized as
\begin{subequations}\label{eq:problem1}
	\begin{align}
	\cP_1:\quad\underset{\{\Qk\},\bPhi,\bXi} \max \quad &\etaSE(\bQ, \bPhi, \bXi), \label{p1a}\\
	{\mathrm{s.t.}} \quad &\tr{\Qk}\leq\Pmaxk,\quad \Qk \succeq \mathbf{0}, \label{p1b}\\
	&\tr{\Rki\Qk}\leq D_{k,i},\quad \forall k,i, \label{p1c}\\
	&\phi_n \in \mathcal{F}_1,\quad \forall n, \label{p1d}\\
	&\bXi \in \mathcal{F}_3^{S\times M}. \label{p1e}
	\end{align}
\end{subequations}
The constraints \eqref{p1b} and \eqref{p1c} keep the transmission power and EM exposure level below some specified values, and \eqref{p1d}, \eqref{p1e} prescribe the formats of RIS phase shift matrix $\bPhi$ and DMA weight matrix $\bXi$,  respectively. 

Note that it is difficult to tackle with $\cP_1$ directly. Firstly, due to the introduction of SAR constraints, the optimization of transmission covariance matrix $\{\bQ_k\}_{\forall k}$ can not be transformed into the traditional power allocation problem, which brings challenges in handling $\bQ_k$. Secondly, calculating \eqref{p1a} in the scenario of partial CSI will bring expensive computational cost. Thirdly, the non-convex constraints of \eqref{p1d} and \eqref{p1e} further complicate the joint optimization problem. With these in mind, we intend to investigate approaches to address these difficulties.

\section{EM Aware SE Optimization with Full CSI} \label{sec:opt_SE_design_PCSI}
With full CSI of $\bH_1$ and $\{\Htwok\}$, the expectation operation in \eqref{equ:SE_model} is removed. Note that the variables of \eqref{p1a} are tightly coupled in $\cP_1$. It is complicated to optimize $\bQ$, $\bPhi$, and $\bXi$ jointly, especially for the case with high dimensional matrix and non-convex constraints. To reduce complexity, we apply the AO method, which is commonly used for problems with numerous optimization variables \cite{YXN21,YXA21}, to optimize $\bQ$, $\bPhi$, and $\bXi$ separately so that the problem can be handled in an alternating manner.


\subsection{Transmit Covariance Matrices Design}\label{subsec:Qk1}
Suppose $\bPhi$ and $\bXi$ are fixed values in the feasible set that satisfy the corresponding constraints. Then \eqref{p1d} and \eqref{p1e} can be regarded as irrelevant constraints when independently optimizing the transmit covariance matrices. 
By applying Sylvester’s determinant theorem \cite{Matrix}, we rewrite \eqref{equ:SE_model} as
\begin{align}\label{equ:SE_model1}
&\etaSE(\Q)=\log\det\Bigg(\bI_{S}+ \ntb
&\left.\frac{1}{\sigmatwo}\sumkeqtoK\bH_1\bPhi\Htwok\Qk\Htwok^H\bPhi^H\bH_1^H\bXi^H(\bXi\bXi^H)^{-1}\bXi\right),
\end{align}
where $\bXi^H(\bXi\bXi^H)^{-1}\bXi$ is the projection matrix \cite[Ch. 5.9]{Matrix}. Denote the compact singular valued decomposition of $\bXi$ as
\begin{align}\label{equ:Xidecompose}
	\bXi=\bU_1\Xitilde\Vtildeone^H,
\end{align}	
where $\bU_1\in\C^{S\times S}$ is the unitary matrix, $\Xitilde\in\C^{S\times S}$ is the diagonal matrix with singular values in the descending order, and $\Vtildeone\in\C^{M\times S}$ is the matrix consisting of the first $S$ columns, which are mutually orthonormal, of the right singular vector matrix of $\bXi$. According to \cite{Matrix}, the projection matrix can be simplified as $\bXi^H(\bXi\bXi^H)^{-1}\bXi=\Vtildeone\Vtildeone^H$. 
Then, \eqref{equ:SE_model1} is equivalent to 
\begin{align}\label{equ:SE_model2}
\etaSE(\Q)&=\log\det\Bigg(\bI_{S}+\ntb
&\left.\frac{1}{\sigmatwo}\sumkeqtoK\Vtildeone^H\bH_1\bPhi\Htwok\Qk\Htwok^H\bPhi^H\bH_1^H\Vtildeone\right).
\end{align}
Denote $\bG_k=\Vtildeone^H\bH_1\bPhi\Htwok\in\C^{S\times N_k}$ as the equivalent channel matrix, then $\cP_1$ for the optimization design of $\{\Qk\}$ with full CSI can be formulated as
\begin{subequations}\label{eq:problem2a}
	\begin{align}
	\cP_{2}^{a}:\ \underset{\{\Qk\}_{\forall k}} \max\  &\log\det\left(\bI_{S}+\frac{1}{\sigmatwo}\sumkeqtoK\bG_k\Qk\bG_k^H\right), \label{p2aa}\\
	{\mathrm{s.t.}} \ &\tr{\Qk}\leq\Pmaxk,\quad \Qk \succeq \mathbf{0}, \label{p2ab}\\
	&\tr{\Rki\Qk}\leq D_{k,i},\quad \forall k,i. \label{p2ac}
	\end{align}
\end{subequations}

It is noteworthy that \eqref{p2aa} is concave on $\{\Qk\}_{\forall k}$, and \eqref{p2ab}, \eqref{p2ac} are linear constraints. Then, $\cP_{2}^{a}$ is a semidefinite programming problem. To this end, we consider its Lagrange dual function written as
\begin{align}\label{equ:Lagrangian1}
&\mathcal{L}\left(\Q,\{\muk\},\{\lambdaki\}\right)
=\etaSE(\bQ) - \sumkeqtoK\muk(\tr{\Qk}-\Pmaxk)\ntb 
&\quad\quad\ -\sumkeqtoK\sumieqtoAk\lambdaki(\tr{\Rki\Qk}-D_{k,i}), 
\end{align} 
where $\muk\geq 0, \lambdaki\geq 0, \forall k,i$ are Lagrange multipliers. As elaborated in \cite{Convex}, the strong dual problem of $\cP_{2}^{a}$ can be expressed as
\begin{align}\label{equ:dual_problem1}
\underset{\{\muk\}_{\forall k},\{\lambdaki\}_{\forall k,i}}\min\ \underset{\{\Qk\}_{\forall k}}\max \quad \mathcal{L}\left(\Q,\{\muk\},\{\lambdaki\}\right).
\end{align}

\begin{prop}\label{Prop:Qk_optimal1}
	Assume the optimal dual values for power and SAR constraints of the problem \eqref{equ:dual_problem1} are $\{\mukopt\}_{\forall k}$ and $\{\lambdakiopt\}_{\forall k,i}$, respectively. Denote $\Kkopt=\mukopt\I_{N_k}+\sumieqtoAk\lambdakiopt\Rki$ and $\bC_{ipn}=\sigmatwo\bI_{S}+\sum\limits_{j=1\atop j\neq k}^{K}\bG_j\Qkopt\bG_j^H$. Then, the optimal $\Qk$ can be expressed by
	\begin{align}\label{Optimal_Q1}
	\Qkopt=\Kkopt^{-1/2}\Uk\Lambdak\Uk^H\Kkopt^{-1/2},
	\end{align}
	where $\Uk$ is the eigenvector matrix derived by
	\begin{align}\label{Kk_Hk_eig_decomp}
	\Kkopt^{-1/2}\Gk^H\bC_{ipn}^{-1}\bG_k\Kkopt^{-1/2}
	=\bU_k\bSigma_k\bU_k^H.	
	\end{align}
	Assume that $\bSigma_k=\mathrm{diag}\{p_{k,1},...,p_{k,N_k}\}$ and $p_{k,1}\geq p_{k,2}\geq ...\geq p_{k,N_k}$ for all $k\in\mathcal{K}$. Then, the optimal power allocation matrix is obtained by
	\begin{align}\label{equ:Lambdak}
	\Lambdak=\mathrm{diag}\{\left(1-1/p_{k,1}\right)^+,...,\left(1-1/p_{k,N_k}\right)^+\}.	
	\end{align}
\end{prop}
The proof of \emph{\propref{Prop:Qk_optimal1}} is similar to that in \cite[Th. 3.6]{YLH15}, thus is omitted in this paper. The EM aware transmit covariance matrices optimization with $\bPhi$ and $\bXi$ fixed is detailed in \textbf{\alref{alg:waterfilling_PCSI}}.
\begin{algorithm}[h]
	\caption{Covariance Matrices Design with Full CSI}
	\label{alg:waterfilling_PCSI}
	\begin{algorithmic}[1]
		\State Initialize $\mu_{k}^{(0)}$, $\lambda_{k,i}^{(0)},\forall k,i$, iteration index $\iota=0$.
		\Repeat
		\State Calculate $\bU_k^{(\iota)}$ and $\Lambdak^{(\iota)},\forall k$ by \eqref{Kk_Hk_eig_decomp} and \eqref{equ:Lambdak}.
		\State Obtain $\Qk^{(\iota+1)},\forall k$ by \eqref{Optimal_Q1}.
		\State Set $\iota=\iota+1$.
		\State Update $\mu_{k}^{(\iota)}$, $\lambda_{k,i}^{(\iota)},\forall k,i$ by minimizing $\mathcal{L}$ in \eqref{equ:dual_problem1}.
		\Until $\{\muk\}_{\forall k}$, $\{\lambdaki\}_{\forall k,i}$ converge.
	\end{algorithmic}
\end{algorithm}  


\subsection{RIS Phase Shift Matrix Design}\label{subsec:bPhi1}
Consider the optimization of variable $\bPhi$ with $\bQ$ and $\bXi$ being fixed. Based on the process similar to the previous section, problem $\cP_1$ regarding the optimization target $\bPhi$ is formulated as
\begin{subequations}\label{eq:problem2b}
 	\begin{align}
 	\cP_{2}^{b}:\underset{\bPhi} \max\ &\log\det\left(\bI_{S}+\frac{1}{\sigmatwo}\Vtildeone^H\bH_1\bPhi\bP\bPhi^H\bH_1^H\Vtildeone\right), \label{p2ba}\\
 	{\mathrm{s.t.}} \ &\phi_n \in \mathcal{F}_1,\quad \forall n, \label{p2bb}
 	\end{align}
\end{subequations}
where $\bP=\sumkeqtoK\Htwok\Qk\Htwok^H$. Because of the non-convexity on both the objective function \eqref{p2ba} and the constraint \eqref{p2bb}, there exist great challenges to deal with $\cP_{2}^{b}$ directly. Actually, it can be observed that \eqref{p2ba} can be regarded as the SE expression of an equivalent communication system with the channel matrix $\Vtildeone^H\bH_1\bPhi\bP^{1/2}$. Specifically, the hypothetical input-output system is expressed as
\begin{align}\label{equ:eqsystem}
\by_e=\Vtildeone^H\bH_1\bPhi\bP^{1/2}\bx_e+\bn_e \in \C^{S\times1},
\end{align}
where $\bx_e\sim\mathcal{CN}(0,\bI_{N_R})$, $\by_e$ and $\bn_e\sim\mathcal{CN}(0,\sigmatwo\bI_{S})$ are the equivalent transmit signal, receive signal and thermal noise, respectively. Denote $\bU_e\in\C^{S\times N_R}$ as the equivalent receiving matrix, the MMSE matrix of the received signal after passing the linear decoder is 
\begin{align}\label{equ:2MSE}
&\bE_e=\mathbb{E}\left\{\left(\bU_e^H\by_e-\bx_e\right)\left(\bU_e^H\by_e-\bx_e\right)^H\right\}=\sigmatwo\bU_e^H\bU_e+\ntb
&(\bU_e^H\Vtildeone^H\bH_1\bPhi\bP^{\frac{1}{2}}-\bI_{N_R})(\bU_e^H\Vtildeone^H\bH_1\bPhi\bP^{\frac{1}{2}}-\bI_{N_R})^H.
\end{align}
According to the weighted MMSE (WMMSE) method elaborated in \cite[Th. 1]{SRL11}, $\cP_{2}^{b}$ is equivalent to a WMMSE minimization problem formulated by
\begin{subequations}\label{eq:problem2bMMSE}
	\begin{align}
	\cP_{2}^{\mathrm{MSE}}:\ \underset{\bW_e,\bU_e,\bPhi} \min\  &h(\bW_e,\bU_e,\bPhi)\triangleq\tr{\bW_e\bE_e}\ntb
	&\qquad\qquad-\log\det(\bW_e), \label{pMSEa}\\
	{\mathrm{s.t.}} \quad &\phi_n \in \mathcal{F}_1,\quad \forall n, \label{pMSEb}
	\end{align}
\end{subequations}
where $\bW_e\in\C^{N_R\times N_R}$ is employed as the auxiliary variable. Note that \eqref{pMSEa} is convex on each optimization variable with the other two variables fixed. To this end, we adopt the BCD method \cite{WGS12}, which is the generalization of AO, to minimize \eqref{pMSEa} by iteratively updating $\bW_e$, $\bU_e$ and $\bPhi$ until converging. When other variables are fixed, it is clear to obtain the closed form solutions of the optimal $\bW_{e}$ and $\bU_e$, which are expressed by
\begin{align}
\label{equ:2Weopt}
&\bW_{e}^{\mathrm{opt}}=\bE_e^{-1},\\
\label{equ:2Ueopt}
&\bU_{e}^{\mathrm{opt}}=(\sigmatwo\bI_S+\Vtildeone^H\bH_1\bPhi\bP\bPhi^H\bH_1^H\Vtildeone)^{-1}\Vtildeone^H\bH_1\bPhi\bP^{\frac{1}{2}}.
\end{align}
With given $\bW_{e}$ and $\bU_e$, $\cP_{2}^{\mathrm{MSE}}$ is reduced to 
\begin{subequations}\label{eq:problem2bMSEPhi}
	\begin{align}
	\cP_{2,\bPhi}^{\mathrm{MSE}}:\ \underset{\bPhi} \min\  &\tr{\bPhi^H\bA\bPhi\bP}-\tr{\bPhi^H\bB^H}-\tr{\bPhi\bB}, \label{pMSEpa}\\
	{\mathrm{s.t.}}\ &\phi_n \in \mathcal{F}_1,\quad \forall n, \label{pMSEpb}
	\end{align}
\end{subequations}
where $\bA=\bH_1^H\Vtildeone\bU_e\bW_e\bU_e^H\Vtildeone^H\bH_1\in\C^{N_R\times N_R}$, and $\bB=\bP^{\frac{1}{2}}\bW_e\bU_e^H\Vtildeone^H\bH_1\in\C^{N_R\times N_R}$. Considering that $\bPhi$ is a diagonal matrix where the modulus of each element is unity, we denote diagonal element vectors $\bphi=[\phi_1,...,\phi_n]^T$, and $\bb=\big[\bB_{1,1},...,\bB_{N_R,N_R}\big]^T$. Then, by employing matrix identity derivation, we get \cite{Matrix}
\begin{align}
\label{equ:identity1}
&\tr{\bPhi^H\bA\bPhi\bP}=\bphi^H(\bA\odot\bP^T)\bphi,\\
\label{equ:identity23}
&\tr{\bPhi^H\bB^H}=\bb^H\bphi^*,\qquad \tr{\bPhi\bB}=\bphi^T\bb.
\end{align}
Therefore, the equivalent problem of $\cP_{2,\bPhi}^{\mathrm{MSE}}$ can be written as
\begin{subequations}\label{eq:problem2bMSEPhiEQ}
	\begin{align}
	\overline{\cP}_{2,\bphi}^{\mathrm{MSE}}:\ \underset{\bphi} \min\  &g(\bphi)=\bphi^H(\bA\odot\bP^T)\bphi-2\Re\{\bphi^H\bb^*\}, \label{pMSEeqa}\\
	{\mathrm{s.t.}}\ &\phi_n \in \mathcal{F}_1,\quad \forall n. \label{pMSEeqb}
	\end{align}
\end{subequations}
\begin{prop}\label{Prop:bPhi_optimal2}
	Denote $\bDelta=\bA\odot\bP^T$ and its maximum eigenvalue as $\lambdamax$. The suboptimal solution of $\overline{\cP}_{2,\bphi}^{\mathrm{MSE}}$ can be obtained by the iterative MM procedure, where each surrogate subproblem constructed from the previous iteration result can be written as
	\begin{subequations}\label{eq:problem2MM}
		\begin{align}
		\overline{\cP}_{2,\bphi}^{(\zeta)}:\ \bphi^{(\zeta+1)}=\argmax{\bphi}\  &G(\bphi|\bphi^{(\zeta)})=\Re\{\bphi^H\bc^{(\zeta)}\}, \label{pMMa}\\
		{\mathrm{s.t.}}\ &|\phi_n|=1\quad n=1,...,N_R, \label{pMMb}
		\end{align}
	\end{subequations}
	where
	\begin{align}\label{equ:iteration_cl}
	\bc^{(\zeta)}=(\lambdamax\bI_{N_R}-\bDelta)\bphi^{(\zeta)}+\bb^*,
	\end{align}
	and $\zeta\in\mathbb{N}$ is the iteration index of MM method. Recover $\bPhi^{(\zeta)}=\mathrm{diag}\{\phi_1^{(\zeta)},...,\phi_{N_R}^{(\zeta)}\}$, then $\left\{\bPhi^{(\zeta)}\right\}^{\infty}_{\zeta=0}$ will converge to the suboptimal solution of the problem \eqref{eq:problem2bMSEPhi} where each point represents the local minimizer of the problem. 
\end{prop}
\IEEEproof Refer to \appref{app:A}.

Please notice that if we express $c_n^{(\zeta)}=\left|c_n^{(\zeta)}\right|\emath^{\jmath\vartheta_n^{(\zeta)}}$ as the $n$th element of $\bc^{(\zeta)}$, it is obvious to obtain the closed form solution of $\bPhi^{(\zeta)}$ where 
\begin{align}\label{equ:optimal_phi}
&\phi_{n,\mathrm{opt}}^{(\zeta+1)}=\emath^{\jmath\vartheta_n^{(\zeta)}}.
\end{align} 
Therefore, the RIS phase shift matrix optimization with $\bQ$ and $\bXi$ fixed is proposed in \textbf{\alref{alg:WMMSE_PCSI}}.
\begin{algorithm}[h]
	\caption{RIS Phase Shift Matrix Design with Full CSI}
	\label{alg:WMMSE_PCSI}
	\begin{algorithmic}[1]
		\State Initialize $\bPhi^{(0)},\bU_e^{(0)},\bW_e^{(0)}$, iteration indices $t=0$, $\zeta$, thresholds $\epsilon_1$, $\epsilon_2$, and calculate $h^{(0)}$ by \eqref{pMSEa}.
		\Repeat
		\State Obtain $\bW_e^{(t+1)}$ and $\bU_e^{(t+1)}$ by \eqref{equ:2Weopt} and \eqref{equ:2Ueopt}.
		\State Set $\zeta=0$, $\bPhi^{(0)}=\bPhi^{(t)}$, and obtain $g(\bphi^{(0)})$ by \eqref{pMSEeqa}.
		\Repeat 
		\State Calculate $\bc^{(\zeta)}$ by \eqref{equ:iteration_cl} and express its phase.
		\State Update $\bphi^{(\zeta+1)}$ with the element obtained by \eqref{equ:optimal_phi}.
		\State Set $\zeta=\zeta+1$.
		\State Calculate $g(\bphi^{(\zeta)})$ by \eqref{pMSEeqa}.
		\Until $|g(\bphi^{(\zeta)})-g(\bphi^{(\zeta-1)})|\leq \epsilon_1$.
		\State Return $\bPhi^{(t+1)}$ with the element $\phi_n^{(\zeta)}$.
		\State Set $t=t+1$.
		\State Update $h^{(t)}$ by \eqref{pMSEa}.
		\Until $|h^{(t)}-h^{(t-1)}|\leq \epsilon_2$.
	\end{algorithmic}
\end{algorithm}

\subsection{DMA Weight Matrix Design}\label{subsec:bXi1}
We suppose both $\bQ$ and $\bPhi$ are fixed when optimizing the weight matrix $\bXi$. Therefore, $\cP_1$ is degenerate into 
\begin{subequations}\label{eq:problem2c}
	\begin{align}
	\cP_{2}^{c}:\quad\underset{\bXi} \max\quad  &\etaSE(\bXi), \label{p2ca}\\
	{\mathrm{s.t.}} \quad &\bXi \in \mathcal{F}_3^{S\times M}. \label{p2cb}
	\end{align}
\end{subequations}
Due to the difficulty caused by the non-convexity of constraints, we start from the unconstrained problem as
\begin{align}\label{eq:problem2cuncons}
\cP_{2}^{ucst}:\ \underset{\Vtildeone} \max\ \log\det\left(\bI_{S}+\Vtildeone^H\bS\Vtildeone\right),
\end{align}
where $\bS=\frac{1}{\sigmatwo}\sumkeqtoK\bH_1\bPhi\Htwok\Qk\Htwok^H\bPhi^H\bH_1^H\in\C^{M\times M}$, and $\Vtildeone$ is denoted in \subsecref{subsec:Qk1}. For problem $\cP_{2}^{ucst}$, \cite[Corollary 2]{SDE19} indicates the closed form of the optimal solution. Denote the eigenvalue decomposition $\bS=\bV\widetilde{\bS}\bV^H$ where the diagonal element of $\widetilde{\bS}$ is arranged in the decending order, then $\widetilde{\bV}_{1,\mathrm{opt}}$ can achieve maximum sum rate of \eqref{eq:problem2cuncons} when $\Vtildeone$ is the first $S$ columns of $\bV$, i.e.,
\begin{align}\label{equ:optimalV1}
\widetilde{\bV}_{1,\mathrm{opt}}=\bV_{[1:S]}.
\end{align}
Substituting \eqref{equ:optimalV1} into \eqref{equ:Xidecompose}, we derive the optimal $\bXi$ without the constraint. It is noteworthy that this result is independent of $\bU_1$ and $\Xitilde$ in \eqref{equ:Xidecompose}, thus arbitrary values of $\bU_1$ and $\Xitilde$ exert no effect on the unconstrained maximal SE. With this in mind, we intend to approximate the optimal weight matrix $\bXi$ satisfying the constraints by reasonably configuring these two values, which is formulated by
\begin{subequations}\label{problem2ccons}
	\begin{align}
	\cP_{2}^{cst}:\ &\underset{\bXi,\bU_1,\Xitilde} \min\  \big|\big|\bXi-\bU_1\Xitilde\Vtildeone^H\big|\big|_\mathrm{F}^2, \label{p2csta}\\
	{\mathrm{s.t.}} \quad &\bXi \in \mathcal{F}_3^{S\times M}, \bU_1\in\mathcal{U}^S, \Xitilde\in\mathcal{D}^S, \label{p2cstb}
	\end{align}
\end{subequations}
where $\mathcal{U}^S$ and $\mathcal{D}^S$ are the sets of unitary matrices and diagonal matrices, respectively, on the dimension of $S\times S$. Resorting to the AO method, we address $\cP_{2}^{cst}$ by iteratively optimize three variables with the other two fixed as follows:

First, as $\bU_1$ and $\Xitilde$ are fixed, it is obvious that the optimal $\bXi$ of \eqref{problem2ccons} is dependent on $\bT=\bU_1\Xitilde\Vtildeone^H$, specifically,
\begin{align}\label{equ:Xi2opt}
&\bXi_{s_1,(s_2-1)L+l}^{\mathrm{opt}}=\ntb
&\left\{
\begin{array}{cccc}
\argmin{\xi_{s_1,l}\in\mathcal{F}_2}\left|\xi_{s_1,l}-\bT_{s_1,(s_2-1)L+l}\right|^2  &s_1=s_2 \\
0 &s_1\neq s_2 \\
\end{array}
\right..
\end{align}
Second, denote $\bT_1=\Xitilde\Vtildeone^H$. As $\bXi$ and $\Xitilde$ are fixed, we assume the left and right singular vector matrices of $\bXi\bT_1^H$ are $\bU_S$ and $\bV_S$, repectively. Then the optimal $\bU_1$ is the solution of Procrustes problem \cite[Ch. 7.4.5]{MatAna}:
\begin{align}\label{equ:U12opt}
&\bU_{1,\mathrm{opt}}=\argmin{\bU_1\in\mathcal{U}^S}\big|\big|\bXi-\bU_1\bT_1\big|\big|_\mathrm{F}^2=\bU_S\bV_S^H.
\end{align}
Third, as $\bXi$ and $\bU_1$ are fixed, we denote $\bT_2=\bXi^H\bU_1=[\bt_{2,1},...\bt_{2,S}]$, $\Vtildeone=[\widetilde{\bv}_{1,1},...,\widetilde{\bv}_{1,S}]$ for the equivalent problem
\begin{align}\label{equ:Xitildeprob}
&\Xitilde_{\mathrm{opt}}=\argmin{\Xitilde\in\mathcal{D}^S}\big|\big|\bT_2^H-\Xitilde\Vtildeone^H\big|\big|_\mathrm{F}^2.
\end{align}
Then, the diagonal entries of optimal $\Xitilde$ can be written as
\begin{align}\label{equ:Xitilde2opt}
&\Xitilde_{s,s}^{\mathrm{opt}}=\max\left\{\frac{\Re\left\{\bt_{2,s}^H\widetilde{\bv}_{1,s}\right\}}{\left|\left|\widetilde{\bv}_{1,s}\right|\right|_\mathrm{F}^2},\delta\right\},\ \forall s\in\{1,...,S\},
\end{align}
where $\delta$ is an infinitesimal positive number \cite[Lemma 2]{SDE19}. The DMA weight matrix optimization with $\bQ$ and $\bPhi$ fixed is detailed in \textbf{\alref{alg:AO_DMAs}}.
\begin{algorithm}[h]
	\caption{DMA Weight Matrix Design with Full CSI}
	\label{alg:AO_DMAs}
	\begin{algorithmic}[1]
		\State Initialize $\bU_1^{(0)},\Xitilde^{(0)}$, iteration index $p=0$, thresholds $\epsilon_3$.
		\State Calculate $\widetilde{\bV}_{1,\mathrm{opt}}$ by \eqref{equ:optimalV1} and then get $\bXi^{(0)}$ by \eqref{equ:Xi2opt}.
		\Repeat
		\State Obtain $\bU_{1}^{(p)}$ by \eqref{equ:U12opt}.
		\State Obtain the diagonal entries of $\Xitilde^{(p)}$ by \eqref{equ:Xitilde2opt}.
		\State Set $p=p+1$.
		\State Update the entries of $\bXi^{(p)}$ by \eqref{equ:Xi2opt}.
		\Until $\left|\left|\bXi^{(p)}-\bXi^{(p-1)}\right|\right|_\mathrm{F}^2\leq \epsilon_3$.
	\end{algorithmic}
\end{algorithm}

\section{EM Aware SE Optimization with Partial CSI} \label{sec:opt_SE_design_SCSI}
As for the RIS-aided uplink, it is often the case that users are moving at high speed while RIS and BS are static in the system. Thus the rapid time-varying of the transmission from users to RIS is nonnegligible, which requires more accurate channel estimation and more frequent updates of the transmission design \cite{WMJ13}. To this end, we assume that the RIS-to-DMAs channel $\bH_1$ is perfectly known while only statistical CSI of the users-to-RIS channels $\{\Htwok\}_{\forall k}$ is available.

As presented in \cite{GJL09}, $\Htwok$ can be decomposed with the adoption of Weichselberger’s model as
\begin{align}\label{equ:decomHtk}
\Htwok=\bU_{2,k}\Htwoktilde\bV_{2,k}^H\in\C^{N_R\times N_k},
\end{align}
where $\bU_{2,k}\in\C^{N_R\times N_R}$ and $\bV_{2,k}\in\C^{N_k\times N_k}$ are deterministic unitary matrices, and $\Htwoktilde\in\C^{N_R\times N_k}$ is a random matrix with elements obeying zero-mean independent distribution. The statistics of CSI is presented by the eigenmode coupling matrix \cite{WJW11}, i.e.,
\begin{align}\label{equ:OmegaHtwok}
\Omegatwok=\mathbb{E}\left\{\Htwoktilde\odot\Htwoktilde^H\right\}\in \R^{N_R\times N_k}.
\end{align}
In addition, \eqref{equ:SE_model} indicates the ergodic SE formulated by
\begin{align}\label{equ:SE_model3}
&\etaSE(\Q, \bPhi, \Vtildeone)=\mathbb{E}_{\{\Htwok\}}\Bigg\{\log\det\Bigg(\bI_{S}+\ntb
&\qquad\quad\left.\left.\frac{1}{\sigmatwo}\sumkeqtoK\Vtildeone^H\bH_1\bPhi\Htwok\Qk\Htwok^H\bPhi^H\bH_1^H\Vtildeone\right)\right\}.
\end{align}
Note that the existence of expectations introduces significant computational overhead. Especially for the Monte Carlo method, by which channel states are exhausted for the calculation of the average SE \cite{LGX16}. A low complexity approach for addressing the expectation operations without averaging is the DE method. Via utilizing the large-dimensional random matrix theory, the DE method can provide deterministic approximations of functionals of random matrices, which are asymptotically accurate as the matrix dimensions grow to infinity at a fixed rate \cite{CD11,WJW11}.
In this way, the accurate approximation of \eqref{equ:SE_model3}, which is defined as $\etaSEb$, can be obtained at a low computational level by using $\Omegatwok,\forall k$. Similar to \secref{sec:opt_SE_design_PCSI}, we adopt AO method to iteratively optimize $\bQ$, $\bPhi$ and $\bXi$.

\subsection{Transmit Covariance Matrices Design}\label{subsec:Qk2}

\subsubsection{Deterministic Equivalent Method}
Assume that $\bG_k=\Vtildeone^H\bH_1\bPhi\bU_{2,k}\Htwoktilde\bV_{2,k}^H$, the ergodic SE in \eqref{equ:SE_model3} is equivalent to \eqref{p2aa} with the expectation of $\{\bG_k\}_{\forall k}$, i.e.,
\begin{align}\label{equ:Gkexp}
	\etaSE(\bQ)=\mathbb{E}_{\{\bG_k\}}\log\det\left(\bI_{S}+\frac{1}{\sigmatwo}\sumkeqtoK\bG_k\Qk\bG_k^H\right).
\end{align}
Denote by $\bU_{\mathrm{G}_k^a}=\Vtildeone^H\bH_1\bPhi\bU_{2,k}\in\C^{S\times N_R}$. Then, following closely to the proof of \cite[Prop. 1]{WJW11}, the asymptotic approximation of \eqref{equ:Gkexp} can be expressed by
\begin{align}\label{equ:DEsGk}
&\etaSEb(\Q)=\sumkeqtoK \log\det\left(\I_{N_k}+\frac{1}{\sigmatwo}\Gammak^a\Qk\right)
\ntb
&\quad\quad+\log\det\left(\I_{S}+\sumkeqtoK\Psik^a\right)-\sumkeqtoK(\gammak^a)^T\Omegatwok\psik^a,
\end{align}
where $\gammak^a \triangleq [\gamma_{k,1}^{a},...\gamma_{k,N_R}^{a}]^T$, $\psik^a \triangleq [\psi_{k,1}^{a},...\psi_{k,N_k}^{a}]^T$ are DE parameters. The other two parameters $\Gammak^a$ and $\Psik^a$ are calculated by
\begin{align}
\label{equ:DE_para_Gammak_Gk}
\Gammak^a&=\bV_{2,k}\mathrm{diag}\{\Omegatwok^T\gammak^a\}\bV_{2,k}^H \in \C^{N_k \times N_k},\\
\label{equ:DE_para_Psik_Gk}
\Psik^a&=\bU_{\mathrm{G}_k^a}\mathrm{diag}\{\Omegatwok\psik^a\}\bU_{\mathrm{G}_k^a}^H \in \C^{S \times S}.
\end{align}
Assume that $r\in\{1,...,N_R\}$ and $n_k\in\{1,...,N_k\}$, $\bu_{\mathrm{G}_{k}^a,r}$ and $\bv_{2,n_k}$ represent the $r$th column and $n_k$th column vectors of $\bU_{\mathrm{G}_k^a}$ and $\bV_{2,k}$, respectively, then
\begin{align}
\label{equ:DE_para_gammak_Gk}
\gamma_{k,r}^{a}&=\bu_{\mathrm{G}_{k}^a,r}^H(\I_{S}+\sumkeqtoK\Psik^a)^{-1}\bu_{\mathrm{G}_{k}^a,r},\ \forall k,r,\\
\label{equ:DE_para_psik_Gk}
\psi_{k,n_k}^{a}&=\bv_{2,n_k}^H\Qk(\sigmatwo\I_{N_k}+\Gammak^a\Qk)^{-1}\bv_{2,n_k},\ \forall k,n_k.
\end{align}
With an initial point of $\psik^a$, the DE parameters $\Gammak^a$ and $\Psik^a$ that are used to approximate the ergodic SE can be obtained by cyclically updating $\gammak^a$ and $\psik^a$ through \eqref{equ:DE_para_Gammak_Gk} -- \eqref{equ:DE_para_psik_Gk}, which inspires us to propose an iterative algorithm as shown in \textbf{\alref{alg:DEs}}.
\begin{algorithm}[h]
	\caption{Deterministic Equivalent Algorithm}
	\label{alg:DEs}
	\begin{algorithmic}[1]
		\State Initialize $\{(\psik^{a})^{(0)}\}$, iteration index $q=0$, threshold $\epsilon_4$.
		\For{$k=1$ to $K$}
		\Repeat
		\State Calculate $(\gammak^{a})^{(q+1)}$ by \eqref{equ:DE_para_Psik_Gk} and \eqref{equ:DE_para_gammak_Gk}.
		\State Update $(\psik^{a})^{(q+1)}$ by \eqref{equ:DE_para_Gammak_Gk} and \eqref{equ:DE_para_psik_Gk} with $\{\bQ_k\}_{\forall k}$.
		\State Set $q=q+1$.
		\Until $\left|\left|(\psik^{a})^{(q)}-(\psik^{a})^{(q-1)}\right|\right|^2\leq \epsilon_4$.
		\State Calculate $\Gammak^a$ and $\Psik^a$ by \eqref{equ:DE_para_Gammak_Gk} and \eqref{equ:DE_para_Psik_Gk} with $(\gammak^{a})^{(q)}$ and $(\psik^{a})^{(q)}$.
		\EndFor
	\end{algorithmic}
\end{algorithm}

\subsubsection{EM-Aware Modified Water-filling Algorithm}
Note that $\etaSEb$ in \eqref{equ:DEsGk} is a tight approximation even when $N_k$ and $N_R$ are small \cite{WJW11}. Replacing \eqref{p2aa} by \eqref{equ:DEsGk}, we arrive at the asymptotic SE optimization problem as follows:
\begin{subequations}\label{eq:problem3a}
	\begin{align}
	\cP_3^a:\quad\underset{\{\Qk\}} \max \quad &\etaSEb(\bQ), \label{p3a}\\
	{\mathrm{s.t.}} \quad &\tr{\Qk}\leq\Pmaxk,\quad \Qk \succeq \mathbf{0}, \label{p3b}\\
	&\tr{\Rki\Qk}\leq D_{k,i},\quad \forall k,i. \label{p3c}
	\end{align}
\end{subequations}
Similar to \emph{\propref{Prop:Qk_optimal1}}, the optimal $\bQ_k$ with partial CSI is also available by a modified water-filling algorithm.
\begin{prop}\label{Prop:Qk_optimal2}
	Assume $\{\mukopt\}_{\forall k}$ and $\{\lambdakiopt\}_{\forall k,i}$ are optimal dual values of power and SAR constraints, respectively, and $\Kkopt=\mukopt\I_{N_k}+\sumieqtoAk\lambdakiopt\Rki$. Denote the eigenvalue decomposition 
	\begin{align}\label{Kk_Gammak_eig_decomp}
	\Kkopt^{-1/2}\Gammak^a\Kkopt^{-1/2}=\bU_k\bSigma_k\bU_k^H.	
	\end{align}
	Then, $\Qkopt$ can be obtained by the same method as \eqref{Optimal_Q1} and \eqref{equ:Lambdak} in \emph{\propref{Prop:Qk_optimal1}}.
\end{prop}
\IEEEproof Refer to Appendix B.

Please note that $\Gammak^a$ in \eqref{Kk_Gammak_eig_decomp} also depends on $\{\Qkopt\}_{\forall k}$. We also applies the AO method, i.e., cyclically calculating $\Qkopt$ in \emph{\propref{Prop:Qk_optimal2}} with given $\Gammak^a$, then update $\Gammak^a$ through the newly adjusted $\Qkopt$, which is reflected in steps 4-10 of \textbf{\alref{alg:waterfilling_SCSI}}.
\begin{algorithm}[h]
	\caption{Optimization of $\{\bQ_{k}\}_{\forall k}$ with Partial CSI}
	\label{alg:waterfilling_SCSI}
	\begin{algorithmic}[1]
		\State Initialize dual variables $\mu_{k}^{(0)}$, $\lambda_{k,i}^{(0)},\forall k,i$, feasible $\Qk^{(0)},\forall k$, iteration indices $u_1=0$, $u_2$, threshold $\epsilon_5$. 
		\Repeat
		\State Set $\{\bQ_{k,(0)}^{(u_1)}\}_{\forall k}=\{\Qk^{(u_1)}\}_{\forall k}$, $u_2=0$.
		\Repeat
		\State Calculate DE parameters $\bGamma_{k,(u_2)}^{a}$ and $\bPsi_{k,(u_2)}^{a}, \forall k$ by $\textbf{\alref{alg:DEs}}$ with given $\bQ_{k,(u_2)}^{(u_1)},{\forall k}$.
		\State Obtain $\bU_{k,(u_2)}$ and $\bLambda_{k,(u_2)},\forall k$ by \eqref{Kk_Gammak_eig_decomp} and \eqref{equ:Lambdak}.
		\State Calculate $\bQ_{k,(u_2+1)}^{(u_1)},{\forall k}$ by \eqref{Optimal_Q1}.
		\State Set $u_2=u_2+1$.
		\Until $\left| \etaSEb\left(\bQ_{k,(u_2)}^{(u_1)}\right)-\etaSEb\left(\bQ_{k,(u_2-1)}^{(u_1)}\right) \right| \leq \epsilon_5$.
		\State Return $\{\Qk^{(u_1+1)}\}_{\forall k}=\{\bQ_{k,(u_2)}^{(u_1)}\}_{\forall k}$.
		\State Set $u_1=u_1+1$.
		\State Update $\mu_{k}^{(u_1)}$, $\lambda_{k,i}^{(u_1)},\forall k,i$ by minimizing $\mathcal{L}$ in \eqref{equ:dual_problem1}.
		\Until $\{\muk\}_{\forall k}$, $\{\lambdaki\}_{\forall k,i}$ converge.
	\end{algorithmic}
\end{algorithm}

\subsection{RIS Phase Shift Matrix Design}\label{subsec:bPhi2}
With fixed $\bQ$ and $\bXi$, the asymptotic approximation of SE, represented by $\etaSEb(\bPhi)$, has the same form as \eqref{equ:DEsGk}. In the DE method, we obtain the DE parameter $\Gammak^a$ and $\Psik^a$ with fixed $\gammak^a$ and $\psik^a$, thus only the second term of \eqref{equ:DEsGk} is the function of $\bPhi$ while the other terms are considered as constants in the optimization of $\bPhi$ \cite{YXN21}. Substituting \eqref{equ:DE_para_Psik_Gk} into \eqref{equ:DEsGk}, we express the problem as
\begin{subequations}\label{eq:problem3b}
	\begin{align}
	&\cP_{3}^{b}:\ \underset{\bPhi} \max\ \log\det\Bigg(\bI_{S}+\ntb
	&\  \sumkeqtoK\Vtildeone^H\bH_1\bPhi\bU_{2,k}\mathrm{diag}\{\Omegatwok\psik^a\}\bU_{2,k}^H\bPhi^H\bH_1^H\Vtildeone\Bigg), \label{p3ba}\\
	&\qquad\quad{\mathrm{s.t.}} \ \phi_n \in \mathcal{F}_1,\quad \forall n. \label{p3bb}
	\end{align}
\end{subequations}
We denote by 
\begin{align}\label{equ:Ptilde}
	\widetilde{\bP}=\sigmatwo\sumkeqtoK\bU_{2,k}\mathrm{diag}\{\Omegatwok\psik^a\}\bU_{2,k}^H,
\end{align}
then problem $\cP_{3}^{b}$ is equivalent to $\cP_{2}^{b}$ that just replaces $\bP$ by $\widetilde{\bP}$, i.e.,
\begin{subequations}\label{eq:problem3bplus}
	\begin{align}
	\cP_{3}^{b}:\underset{\bPhi} \max\ &\log\det\left(\bI_{S}+\frac{1}{\sigmatwo}\Vtildeone^H\bH_1\bPhi\widetilde{\bP}\bPhi^H\bH_1^H\Vtildeone\right), \label{p3baa}\\
	{\mathrm{s.t.}} \ &\phi_n \in \mathcal{F}_1,\quad \forall n. \label{p3bbb}
	\end{align}
\end{subequations}
Therefore, the optimal $\bPhi$ with partial CSI will be derived by the same method in \subsecref{subsec:bPhi1}. Above we consider the case where $\psik^a$ is fixed. However, as $\bU_{\mathrm{G}_k^a}$ is one of the matrices for calculating DE parameters, the value of $\psik^a$ is required to be updated along with the variation of $\bPhi$.

\subsection{DMA Weight Matrix Design}
Similar to \subsecref{subsec:bXi1}, we firstly consider the unconstrained problem and utilize the decomposition of \eqref{equ:Xidecompose}. Based on the DE method, the asymptotic approximation of the system SE $\etaSEb(\Vtildeone)$ can also be expressed by \eqref{equ:DEsGk}, where only the second term that contains $\bU_{\mathrm{G}_k^a}$ is related to $\Vtildeone$ while the others are constants. On this condition, we obtain the same objective function as \eqref{p3ba}. Note that when denoting
\begin{align}\label{equ:Stilde}
	\widetilde{\bS}=\bH_1\bPhi\bU_{2,k}\mathrm{diag}\{\Omegatwok\psik^a\}\bU_{2,k}^H\bPhi^H\bH_1^H,
\end{align}
the optimization problem of $\bXi$ without the constraint is expressed by \begin{align}\label{eq:problem3cuncons}
\cP_{3}^{ucst}:\ \underset{\Vtildeone} \max\ \log\det\left(\bI_{S}+\Vtildeone^H\widetilde{\bS}\Vtildeone\right),
\end{align}
which is converted to the same problem form as \eqref{eq:problem2cuncons}. In the following, the optimization process of constrained $\bXi$ is similar to the method in \subsecref{subsec:bXi1}, thus omitted here. Note that when $\psik^a$ is considered unfixed, the value of $\widetilde{\bS}$ is related to $\Vtildeone$, which inspires us to update the value of $\widetilde{\bS}$ in the overall AO method.

\section{Overall Algorithm and Analysis}\label{sec:analysis}
Now that we have studied the optimization of covariance matrices $\{\bQ\}_{\forall k}$, RIS phase shift matrix $\bPhi$ and DMA weight matrix $\bXi$ separately under both full and partial CSI. In this section, we propose the overall EM aware SE maximization algorithm and then analyze its convergence and complexity. Note that whether with full or partial CSI, the overall algorithm shares the same AO-based framework. Therefore, we focus our analysis on the case with partial CSI, which has been studied in \secref{sec:opt_SE_design_SCSI}. Then the convergence performance and the complexity with full CSI can be obtained in a similar way. 

Compared with \secref{sec:opt_SE_design_PCSI} that consider the full CSI case, \secref{sec:opt_SE_design_SCSI} adopts the DE method to approximate the system SE without requiring time-consuming expectation operation. The DE parameters used for asymptotic approximation can be obtained by using iteration in \textbf{\alref{alg:DEs}} until $\gammak^a$ and $\psik^a$ converge to the unique solution point \cite{WJW11}. By using the AO-based framework, the overall transmission strategy with partial CSI is detailed in \textbf{\alref{alg:AOmethod}}. 
Similarly, we get the AO-based algorithm with full CSI i.e., iterating \textbf{Algorithm 1, 2, 3} until the system performance converges.

\begin{algorithm}[h]
	\caption{AO-based Method for SE Maximization with Partial CSI}
	\label{alg:AOmethod}
	\begin{algorithmic}[1]
		\Require Initial $\bQ^{(0)}$, $\bPhi^{(0)}$, $\bXi^{(0)}$, iteration index $\ell$, threshold $\epsilon_6$.
		\Ensure Approximation solutions $\bQ_{\mathrm{opt}}$, $\bPhi_{\mathrm{opt}}$, and $\bXi_{\mathrm{opt}}$.
		\State Set $\ell=0$ and calculate $\etaSEb^{(0)}$ by \eqref{equ:DEsGk}.
		\Repeat 
		\State Obtain $\bQ^{(\ell+1)}=\mathrm{diag}\{\bQ_k^{(\ell+1)}\}_{k=1}^{K}$ by \textbf{\alref{alg:waterfilling_SCSI}}.
		\State Get $\bm{\psi}_{k}^a$ by \textbf{\alref{alg:DEs}} and calculate $\widetilde{\bP}^{(\ell)}$ by \eqref{equ:Ptilde}.
		\State Obtain $\bPhi^{(\ell+1)}$ by \textbf{\alref{alg:WMMSE_PCSI}}.
		\State Get $\bm{\psi}_{k}^a$ by \textbf{\alref{alg:DEs}} and calculate $\widetilde{\bS}^{(\ell)}$ by \eqref{equ:Stilde}.
		\State Obtain $\bXi^{(\ell+1)}$ by \textbf{\alref{alg:AO_DMAs}}.
		\State Update $\ell=\ell+1$.
		\State Calculate $\etaSE^{(\ell)}$ with $\bQ^{(\ell)}$, $\bPhi^{(\ell)}$ and $\bXi^{(\ell)}$ by \eqref{equ:DEsGk}.
		\Until $\left|\etaSEb^{(\ell)}-\etaSEb^{(\ell-1)}\right|\leq \epsilon_6$.
		\State Return $\bQ_{\mathrm{opt}}=\bQ^{(\ell)}$, $\bPhi_{\mathrm{opt}}=\bPhi^{(\ell)}$ and $\bXi_{\mathrm{opt}}=\bXi^{(\ell)}$.
	\end{algorithmic}
\end{algorithm}

In general, \textbf{\alref{alg:AOmethod}} adopts the method that $\bQ$, $\bPhi$ and $\bXi$ are optimized alternatively. Due to the fact that \eqref{eq:problem2a} and \eqref{eq:problem3a} are convex problems, the solutions of \emph{\propref{Prop:Qk_optimal1}} and \emph{\propref{Prop:Qk_optimal2}} are essentially the results of solving their strong dual problems like \eqref{equ:dual_problem1}. Therefore, the water-filling algorithm by iterating dual variables in \textbf{\alref{alg:waterfilling_PCSI}} and \textbf{\alref{alg:waterfilling_SCSI}} will definitely converges to the global optimal solutions of problems \eqref{eq:problem2a} and \eqref{eq:problem3a}, respectively \cite{Convex}. In addition, \textbf{\alref{alg:WMMSE_PCSI}} solve the WMMSE problem in \eqref{eq:problem2bMMSE} by adopting BCD method, of which the convergence is ensured according to \cite{SRL11}. \textbf{\alref{alg:AO_DMAs}} utilizes the AO method for problem \eqref{problem2ccons} with multiple optimization variables. As the object function in \eqref{p2csta} is differentiable over all the variables, the AO algorithm is ensured to converge \cite{BH03}. Therefore, the value of system SE will not reduce through steps 3, 5, 7 in \textbf{\alref{alg:AOmethod}}, which guarantees the convergence of the overall algorithm. Likewise, the convergence under full CSI scenario can be ensured by the fact the optimization objective is upper-bounded and does not decrease after each AO process.

Notice that the complexity of the proposed algorithm is related to the number of iterations, which can be adjusted by varying the input thresholds. As the result of the fast convergence of DE parameters, the complexity of \textbf{\alref{alg:DEs}} can be almost ignored [47]. In \textbf{\alref{alg:waterfilling_SCSI}}, the major complexity of steps 5--8 locates in (50) and can be expressed by $\mathcal{O}(N_k^3)$. Moreover, step 12 updates dual variables through minimizing $\mathcal{L}$, which has the complexity of $\mathcal{O}((K+\sum_{k}A_k)^x)$, where $1\leq x\leq 4$ for the convex program [33]. Suppose the numbers of iterations for the inner AO and the outer water-filling are $I_{\mathrm{A}}$ and $I_{\mathrm{W}}$, respectively. Then, the complexity of \textbf{\alref{alg:waterfilling_SCSI}} is estimated with $\mathcal{O}(I_{\mathrm{W}}(I_{\mathrm{A}}\sum_{k}N_k^3+(K+\sum_{k}A_k)^x))$. As \textbf{\alref{alg:waterfilling_PCSI}} has no need to obtain the optimal DE parameters $\Gammak^a$, its complexity can be witten as $\mathcal{O}(I_{\mathrm{W}}(\sum_{k}N_k^3+(K+\sum_{k}A_k)^x))$. \textbf{\alref{alg:WMMSE_PCSI}} is composed of the outer BCD and the inner MM algorithms, which are assumed to go through $I_{\mathrm{B}}$ and $I_{\mathrm{M}}$ iterations, respectively. Because of the matrix inversion operations in (23) and (24), their complexity are $\mathcal{O}(N_R^3)$ and $\mathcal{O}(S^3)$, respectively. In step 6, the eigenvalues of $\bm{\Delta}$ need to be pre-calculated, which introduces the complexity of $\mathcal{O}(N_R^3)$. In addition, the complexity of MM algorithm mainly centers at the calculation of $\bc^{(\zeta)}$, which is $\mathcal{O}(N_R^2)$. Considering that usually $N_R^3\gg S^3$, the complexity of \textbf{\alref{alg:WMMSE_PCSI}} can be estimated as $\mathcal{O}(I_{\mathrm{B}}(N_R^3+I_{\mathrm{M}}N_R^2))$. The calculation of each iteration in \textbf{\alref{alg:AO_DMAs}} is concentrated in (37), which has the complexity of $\mathcal{O}(S^3)$. Assume the number of iterations of \textbf{\alref{alg:AO_DMAs}} is $I_{\mathrm{O}}$, its complexity is estimated with $\mathcal{O}(I_{\mathrm{O}}S^3)$. Based on the analysis above, the complexity of \textbf{\alref{alg:AOmethod}} is expressed as $\mathcal{O}(I_{\mathrm{AO}}(I_{\mathrm{W}}(I_{\mathrm{A}}\sum_{k}N_k^3+(K+\sum_{k}A_k)^x)+I_{\mathrm{B}}(N_R^3+I_{\mathrm{M}}N_R^2)+I_{\mathrm{O}}S^3)$, where $I_{\mathrm{AO}}$ is the number of iterations of the overall AO. Adopting the same notations, the AO-based algorithm with full CSI has the above complexity form removing the number of iterations, $I_{\mathrm{A}}$.

\section{Numerical Results}\label{sec:numerical_results}
In this section, the simulation results are given to appraise the SE performance of the EM aware transmission strategy in which RIS and DMAs are considered in the multiuser MIMO system. Our simulation focuses on the main contributions of the proposed algorithm, i.e, the SE maximization method under both full CSI and partial CSI. In practical scenarios, $\mathcal{F}_3^{S\times M}$ in \eqref{p1e} is determined by the ability to externally control the response of each element, which comprises of four classical DMA weights \cite{SDE19,WSE21}, i.e.,
\begin{itemize}
	\item[(1).] $\mathrm{{UC}}$: $\mathcal{F}_2=\C$.
	\item[(2).] $\mathrm{{AO}}$: $\mathcal{F}_2=[0.001,2]$.
	\item[(3).] $\mathrm{{BA}}$: $\mathcal{F}_2=\{0,0.1\}$.
	\item[(4).] $\mathrm{{LP}}$:  $\mathcal{F}_2=\left\{\frac{\jmath+\emath^{\jmath\varphi}}{2}|\varphi\in\left[\right.0,2\pi\left.\right)\right\}$.
\end{itemize}
Notice that our simulations are mainly based on unconstrained DMA weights, i.e., $\mathcal{F}_2=\C$, unless explicitly mentioned otherwise. The parameters used in the simulation are listed in \tabref{parameters} \cite{YLH17,XYN21}.
\begin{table}[htbp]
	\caption{Simulation Parameters}\label{parameters}
	\centering
	\ra{1.3}
	\footnotesize
	\begin{tabular}{LR}
		\toprule
		Parameter & Value  \\
		\midrule
		\rowcolor{lightblue}
		Channel model     & 3GPP SCM    \\
		Values of $\bU_{2,k}$ and $\bV_{2,k},\forall k$ & DFT matrices\\
		\rowcolor{lightblue}
		Noise covariance                & -96 dBm  \\
		Path loss      &120 dB                            \\
		\rowcolor{lightblue}
		Number of users &$K=4$\\
		Number of attennas at users & $N_k=4,\forall k$\\
		\rowcolor{lightblue}
		Number of SAR constraints & $A_k=1,\forall k$\\
		Number of RIS phase shift units & $N_R=16$ \\
		\rowcolor{lightblue}
		Number of DMA microstrips & $S=8$\\
		Number of metamaterial units per microstrip & $L=8$\\
		\bottomrule
	\end{tabular}
\end{table}

Without loss of generality, the power constraints and SAR constraints are set to be the same for all users for the clarity of the simulation results \cite{YLH17}. Therefore, we assume that $\Pmaxk=\Pmax, \forall k$ and $D_{k,i}=D=0.8\ \mathrm{W/kg}, \forall k,i$ in our simulation. The SAR matrix corresponding to the SAR budget $D$ is assigned as \cite{YLH15}
\begin{align}
\Rki=\bR=
\left[
\begin{array}{cccc}
8 &-6\jmath &-2.1 &0 \\
6\jmath &8 &-6\jmath &-2.1 \\
-2.1 &6\jmath &8 &-6\jmath \\
0 &-2.1 &6\jmath &8
\end{array}
\right].
\end{align}

\subsection{Convergence Performance}
\begin{figure}[t]
	\centering
	\subfloat[]{\centering\includegraphics[width=0.35\textwidth]{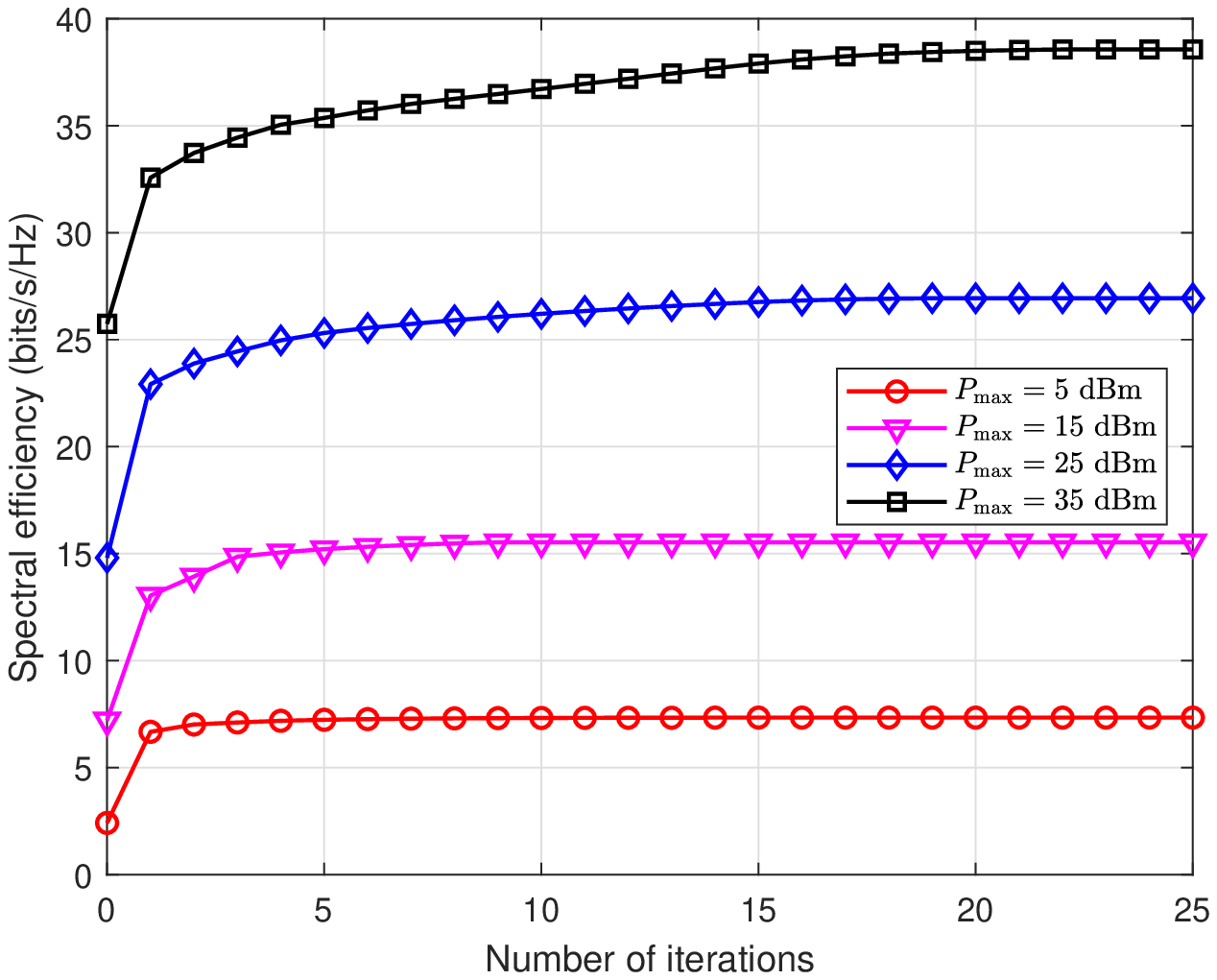}\label{fig:iterICSI}}
	\hfill
	\subfloat[]{\centering\includegraphics[width=0.35\textwidth]{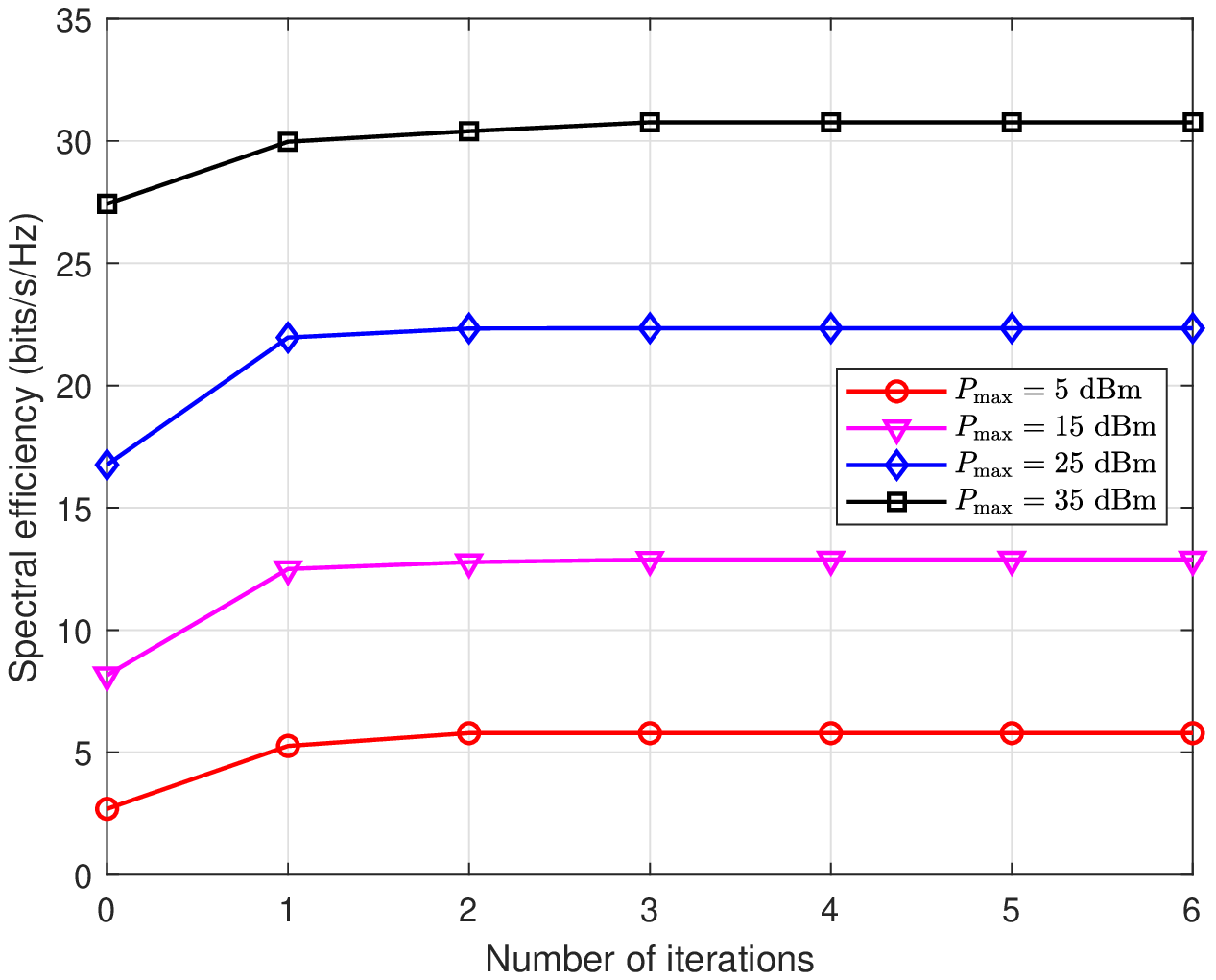}\label{fig:iterSCSI}}
	\caption{Convergence of the AO-based method for EM exposure constrained SE maximization design in hybrid RIS and DMA assisted system. (a) Full CSI;  (b) Partial CSI.}
	\label{fig:iteration}
\end{figure}

\figref{fig:iteration} evaluates the number of iterations required for the proposed EM aware SE maximization method in RIS and DMA assisted system, where we jointly optimize the covariance matrix $\bQ$, RIS phase shift matrix $\bPhi$ and DMA weight matrix $\bXi$ to maximize system SE with the determined values of SAR matrix $\bR$ and SAR budget $D$. Note that \subfigref{fig:iteration}{fig:iterSCSI} presents the iteration state in \textbf{\alref{alg:AOmethod}} and \subfigref{fig:iteration}{fig:iterICSI} corresponds to the overall algorithm with full CSI. 
The results illustrate the rapid convergence rates of the partial CSI scenario in specific power budget regions compared with the full CSI scenario. However, when the perfect CSI of $\Htwok$ is available, the computational complexity of the overall algorithm is greatly reduced, which leads to the fast running speed of the overall algorithm.

\subsection{Impact of EM Exposure and RIS on SE Optimization}
\begin{figure}[t]
	\centering
	\includegraphics[width=0.35\textwidth]{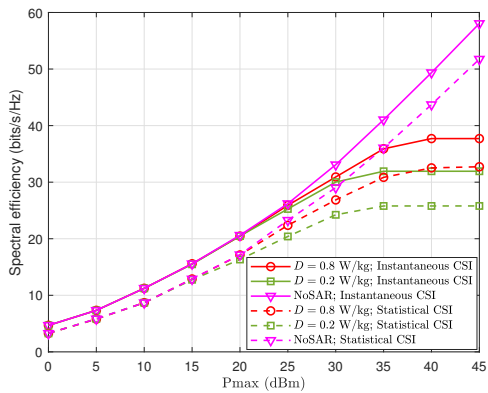}
	\caption{Comparison of SE performance between the optimization with and without EM exposure constraints.}
	\label{fig:sepnoSAR}
\end{figure}

\begin{figure}[t]
	\centering
	\includegraphics[width=0.35\textwidth]{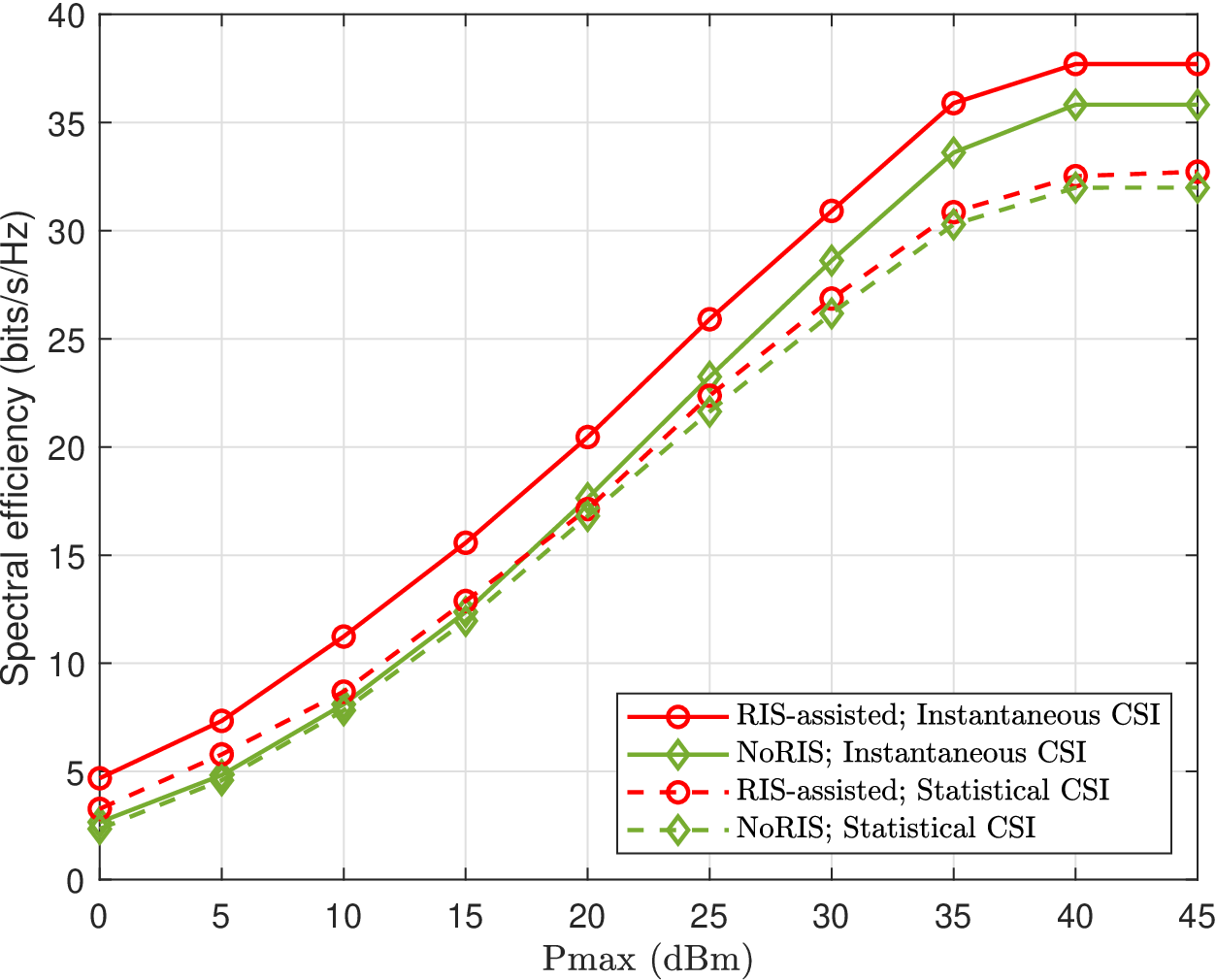}
	\caption{Comparison of SE performance between the RIS assisted system and the non-RIS assisted system.}
	\label{fig:sepUCnoRIS}
\end{figure}

\figref{fig:sepnoSAR} sketches the trend of overall SE performance versus the variation of power budget $\Pmax$ in the presence and absence of EM exposure constraints, respectively. It can be observed that the system SE in our work is not positively correlated with the transmit power as traditional ways, but first increases speedily with the increase of $\Pmax$, then become constant when $\Pmax$ is relatively large. This is mainly due to the EM exposure constraints introduced into the problem, i.e., \eqref{p1c}. When there is a tiny power budget, $\Pmax$ is the main factor limiting the growth of system SE where SAR constraints are easily satisfied even all the power is used. With the increase of $\Pmax$, SAR constraints gradually become the main restriction of the system SE and eventually lead to the saturation of the curves. Actually, this process is affected by the SAR budget $D$. As shown in \figref{fig:sepnoSAR}, the larger SAR budget means more relaxed SAR constraints, which brings higher SE performance and requires more power budget to achieve the saturation of the curves.

\figref{fig:sepUCnoRIS} compares the SE performance of the proposed approach in the RIS assisted system with the non-RIS assisted system when DMAs are configured at the BS. As expected, for both full CSI and partial CSI, optimization without RIS results in the reduction on the system SE compared to that with optimized RIS shift matrix $\bPhi$. In fact, if the $\bPhi_{\mathrm{opt}}$ optimized by \textbf{\alref{alg:AOmethod}} is an identity matrix, the system with or without RIS assist will achieve the same SE. In addition, the optimization of the RIS phase shift matrix is related to the optimal transmission covariance matrix, as shown in steps 4, 5 of \textbf{\alref{alg:AOmethod}}. Therefore, the solution $\bPhi_{\mathrm{opt}}$ under different power budgets will not be unified and require independent optimization, which emphasizes the necessity of joint optimization design.

\subsection{Comparison With Conventional Antennas}
\begin{figure}[t]
	\centering
	\includegraphics[width=0.35\textwidth]{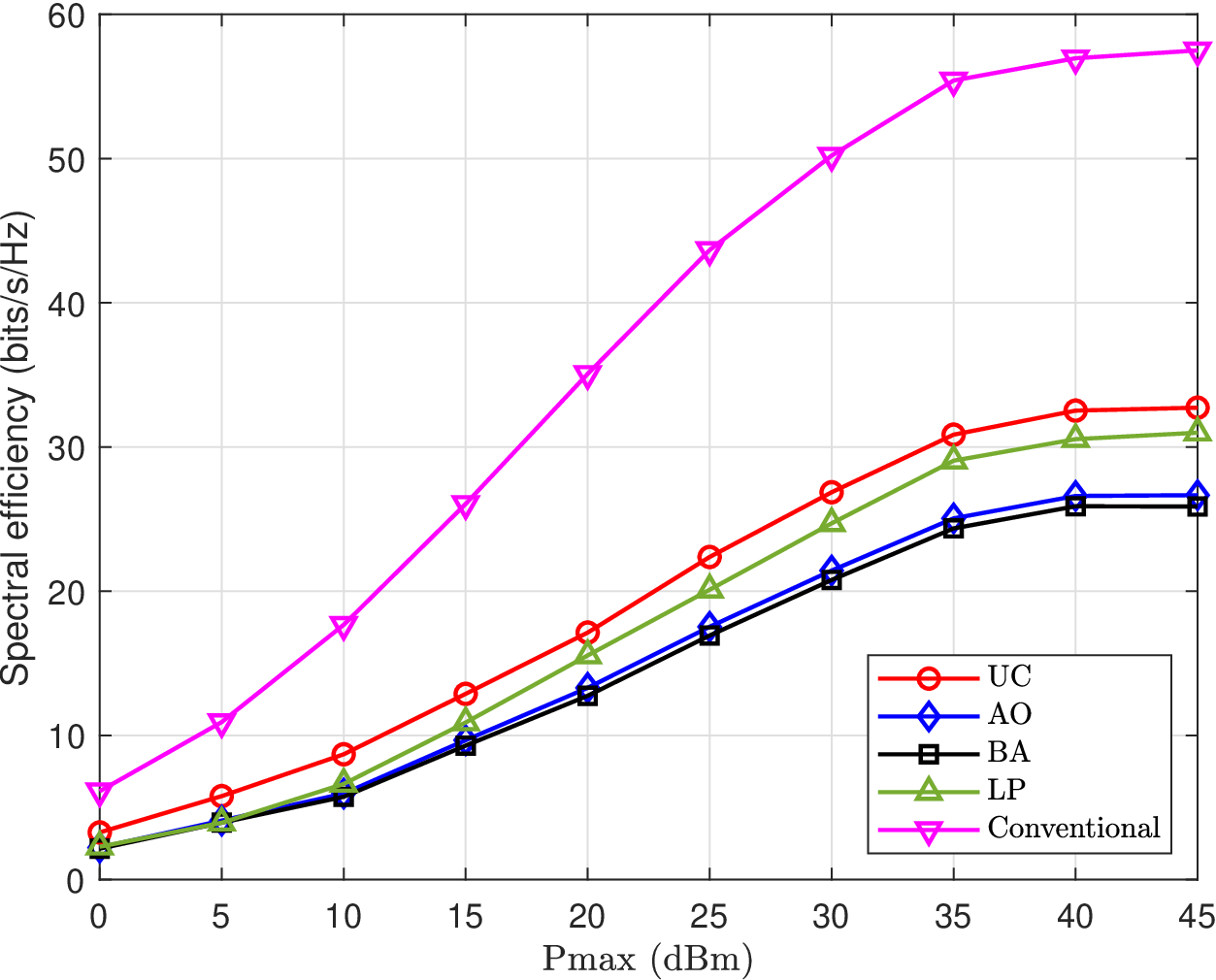}
	\caption{Comparison of system SE between the DMA assisted and conventional antenna assisted system with partial CSI.}
	\label{fig:sepSCSIconventional}
\end{figure}

\figref{fig:sepSCSIconventional} compares the SE performance of two scenarios when the BS is equipped with DMAs and that with conventional antennas. In the conventional system, the BS uses uniform linear arrays (ULAs) with antenna spacing of half-wavelength, where each antenna is associated with an independent RF chain \cite{YXN21}. 
Suppose the same condition that the BS is placed with $M=SL=64$ antennas, which constitute a total of $64$ RF chains, and then the system SE is denoted by
\begin{align}\label{equ:SE_conventional}
&\etaSE=\mathbb{E}_{\{\Htwok\}}\Bigg\{\log\det\Bigg(\bI_{M}+\frac{1}{\sigmatwo}\sumkeqtoK\bH_1\cdot\ntb
&\qquad\qquad\bPhi\Htwok\Qk\Htwok^H\bPhi^H\bH_1^H\Bigg)\Bigg\}\ \bps/\Hz.
\end{align}
It is evident in \figref{fig:sepSCSIconventional} that due to more RF chains at the BS side, the conventional antenna assisted system achieves higher SE compared with the DMA assisted system. However, the significant demand for RF chains tremendously increases the power consumption and the hardware cost, which are commendably reduced by applying DMAs.

In \figref{fig:sepSCSIconventional}, we consider four classical kinds of feasible sets of DMA weights, including unconstrained weights, amplitude only, binary amplitude, and Lorentzian phase. As depicted, system SE is achieved to the maximum when DMA weights range at a complex plane, followed by the SE performance of $\mathrm{SE_{LP}}$, $\mathrm{SE_{AO}}$, and $\mathrm{SE_{BA}}$. It is reasonable because the feasible set of unconstrained DMA weights contains the other three conditions. On the conditions of the Lorentzian phase, the optimal value of DMA weights is searched in a portion of the complex plane, which makes the optimized SE close to the $\mathrm{SE_{UC}}$. By comparison, cases amplitude only and binary amplitude optimize DMA weights only on the one-dimensional real axis, which reduces the upper bound of the system SE and achieves almost the same performance. In practice, implementing DMA with binary amplitude has lower hardware costs, which makes it more attractive than the DMA with amplitude only.

\subsection{Comparison With Baseline Approaches}
To further demonstrate the effectiveness of the active design that considers EM exposure constraints, we compare the system SE of the proposed algorithm with baseline approaches in \figref{fig:sepSCSIconventional}, which contain worst-case power backoff and adaptive backoff approach. Both two baseline approaches only consider the power constraints in $\cP_3^a$, i.e., \eqref{p3c} is not taken into account, and then make the solution in $\cP_3^a$ satisfy the SAR constraints by introducing the backoff factor $\rho$. The specific process of the two baseline approaches is detailed as follows:
\begin{itemize}
	\item \emph{Worst-Case Power Backoff} \cite{YLH15}: Denote the worst-case power backoff factor as 
	\begin{align}\label{equ:WCbackoff2}
	\rho_1=\min\left\{1,D/\mathrm{SAR^{worst}}\right\},
	\end{align}
	where $\mathrm{SAR^{worst}}$ is the worst SAR given by
	\begin{align}\label{equ:WCbackoff3}
	\mathrm{SAR^{worst}}=\underset{k}\max\underset{\tr{\Qk}\leq\Pmaxk}\max \tr{\bR\Qk}.
	\end{align}
	This approach only considers the power constraints and then makes the final result satisfy the SAR constraints by reducing the power budgets as 
	\begin{align}\label{equ:WCbackoff1}
	\bQ^{\mathrm{wc}}_{\mathrm{opt}}=\argmax{\tr{\Qk}\leq\rho_1\Pmaxk}\ \etaSEb(\bQ), \quad \forall k.
	\end{align}

	\item \emph{Adaptive Backoff} \cite{YLH15}: By omitting the SAR constraints, problem $\cP_3^a$ degenerates into
	\begin{align}\label{equ:Adpbackoff1}
	\bQ^{\mathrm{0}}=\argmax{\tr{\Qk}\leq\Pmaxk}\ \etaSEb(\bQ), \quad \forall k.
	\end{align}
	Then, the adaptive backoff factor $\rho_2$ is introduced to make the result meet the SAR constraints.
	\begin{align}\label{equ:Adpbackoff2}
	\rho_{2(k)}=\min\left\{1,D/\tr{\bR\bQ^{\mathrm{0}}_{{k}}}\right\},\quad \forall k.
	\end{align}
	Finally, the corresponding transmit covariance can be expressed as
	\begin{align}\label{equ:Adpbackoff3}
	\bQ^{\mathrm{adp}}_{k,\mathrm{opt}}=\rho_{2(k)}\Qk^{\mathrm{0}},\ \forall k.
	\end{align}
\end{itemize}
It can be observed that when $\Pmax$ is lower than a certain threshold, the three methods achieve the same SE performance. Under this condition, SAR constraints can be naturally satisfied even when the power budget is fully used to transmit signals, which means \eqref{p2ac} or \eqref{p3c} is actually negligible, and all the backoff factors $\rho_1$, $\rho_{2(k)}, \forall k$ are equal to one. When $\Pmax$ is higher than a threshold, SAR constraints start to restrict the increase of system SE in a different way from power constraints. Since the backoff methods only consider the SAR constraint as the additional power constraint and ignore the correlation between SAR and the phase differences of transmit antennas, our proposed method, which considers both the power and SAR constraints, can achieve higher SE.

\begin{figure}[t]
	\centering
	\includegraphics[width=0.35\textwidth]{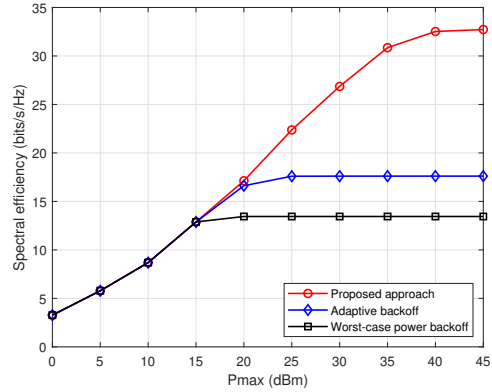}
	\caption{SE performance comparison between the proposed and baseline approaches with partial CSI.}
	\label{fig:sepSCSIbackoff}
\end{figure}

\section{Conclusion}\label{sec:conclusion}
To summarize, we investigated the EM exposure constrained uplink strategy design of the hybrid RIS and DMA assisted multiuser MIMO system. We considered a practical scenario where instantaneous CSI is always available for the RIS-to-DMAs channel and discussed the instantaneous and statistical CSI of users-to RIS channel, respectively. To obtain the maximal SE under full or partial CSI, we proposed the corresponding algorithm to jointly optimize the transmit covariance matrices, RIS phase shift matrix, and DMA weight matrix, where power and SAR constraints were included in the optimization. Especially, for the reduction of the computational complexity under partial CSI, we adopted the DEs to asymptotically approximate the ergodic SE and then proposed a SE maximization algorithm that is applicable for partial CSI of users-to-RIS channels. In the numerical simulations, we verified the convergence performance of the overall algorithm and figured out the impact of EM exposure, RIS and DMA on the system SE by comparing with conventional systems. Then, numerical results substantiated the superior SE performance of the EM aware joint optimization design over the backoff approaches.

\appendices

\section{Proof of \propref{Prop:bPhi_optimal2}}\label{app:A}
According to \cite{MW78}, MM method is an iterative optimization method used to find the approximate solution, where the key is using the local minimizer points of each iteration $\bphi^{(\ell)}$ to construct a series of upper bound functions $\widetilde{g}(\bphi|\bphi^{(\ell)})$ of the original objective $g(\bphi)$ satisfying
\begin{itemize}
	\item[(i),] $\widetilde{g}(\bphi|\bphi^{(\ell)})\geq g(\bphi),\quad \forall \phi_n, \phi_n^{(\ell)}\in\mathcal{F}_1$,
	\item[(ii),] $\widetilde{g}(\bphi^{(\ell)}|\bphi^{(\ell)})= g(\bphi^{(\ell)}),\quad \forall\phi_n^{(\ell)}\in\mathcal{F}_1$,
	\item[(iii),] $\nabla_{\bphi}\widetilde{g}(\bphi^{(\ell)}|\bphi^{(\ell)})= \nabla_{\bphi}g(\bphi^{(\ell)}),\quad \forall\phi_n^{(\ell)}\in\mathcal{F}_1$.	
\end{itemize}
Based on this condition, the sequence $\left\{\widetilde{g}(\bphi^{(\ell)}|\bphi^{(\ell)})\right\}_{\ell=0}^{\infty}$ is monotonically non-increasing and finally converge to the approximation minimum value of $\overline{\cP}_{2,\bphi}^{\mathrm{MSE}}$ with the solution $\bphi_{opt}$ fulfilling the first order Lagrangian conditions. In the following, we prove that the function constructed in \emph{\propref{Prop:bPhi_optimal2}} satisfies the above conditions.

Notice $\bDelta$ is a positive semidefinite matrix as $\bA$ and $\bP$ are both positive semidefinite, so is $\lambdamax\bI_{N_R}-\bDelta$. Applying that $\left|\left|\left(\lambdamax\bI_{N_R}-\bDelta\right)^{\frac{1}{2}}\bphi-\left(\lambdamax\bI_{N_R}-\bDelta\right)^{\frac{1}{2}}\bphi^{(\ell)}\right|\right|^2\geq 0$, i.e.,
\begin{align}\label{equ:appendix1a}
&\bphi^H\left(\lambdamax\bI_{N_R}-\bDelta\right)\bphi+(\bphi^{(\ell)})^H\left(\lambdamax\bI_{N_R}-\bDelta\right)\bphi^{(\ell)}\ntb
&\qquad\qquad-2\Re\left\{\bphi^H(\lambdamax\bI_{N_R}-\bDelta)\bphi^{(\ell)}\right\}\geq 0,
\end{align} 
we obtain the upper bound function
\begin{align}\label{equ:appendix1b}
&g(\bphi)\leq\widetilde{g}(\bphi|\bphi^{(\ell)})=2\lambdamax\bI_{N_R}-(\bphi^{(\ell)})^H\bDelta\bphi^{(\ell)}\ntb
&\quad-2\Re\left\{\bphi^H(\lambdamax\bI_{N_R}-\bDelta)\bphi^{(\ell)}\right\}-2\Re\{\bphi^H\bb^*\}.
\end{align} 
In addition, the identities in conditions (ii) and (iii) also hold referring to \cite{HZA19}. As the properties of MM method, we aims to address the subproblem iteratively as follows:
\begin{subequations}\label{eq:problem2MMdedu}
	\begin{align}
	\overline{\cP}^{(\ell)}:\ \bphi^{(\ell+1)}=\argmin{\bphi}\  &\widetilde{g}(\bphi|\bphi^{(\ell)}), \label{pMMadedu}\\
	{\mathrm{s.t.}}\ &\phi_n \in \mathcal{F}_1,\quad \forall n, \label{pMMbdedu}
	\end{align}
\end{subequations}
which is equivalent to $\overline{\cP}_{2,\bphi}^{(\zeta)}$. This concludes the proof.

\section{Proof of \propref{Prop:Qk_optimal2}}\label{app:B}
By approximating $\etaSE(\bQ)$ as $\etaSEb(\bQ)$ in \eqref{equ:Lagrangian1}, we can obtain the Lagrange dual function of $\cP_3^a$ represented by $\mathcal{L}(\bQ)$. Thus the dual problem is expressed by \eqref{equ:dual_problem1}. As shown in \cite{WJW11}, the derivative of $\etaSEb(\bQ)$ in \eqref{equ:DEsGk} over  $\Qk$ is given by
\begin{align}\label{equ:partialSE}
\frac{\partial\etaSEb(\bQ)}{\partial\Qk}=(\sigmatwo\I_{N_k}+\Gammak^a\Qk)^{-1}\Gammak^a.
\end{align}
Consider the Karush-Kuhn-Tucker (KKT) conditions:
\begin{align}\label{equ:KKT}
\frac{\partial\mathcal{L}}{\partial\Qkopt}=(\sigmatwo\I_{N_k}+\Gammak^a\Qkopt)^{-1}\Gammak^a-\Kkopt=\bm{0}.
\end{align}
Therefore, with fixed $\Gammak^a$ and given $\muk,\ \lambdaki\ \forall k,i$, we can obtain the equivalent inner optimization of \eqref{equ:dual_problem1} as
\begin{align}\label{equ:innermaxL}
\underset{\Qk}\max \ \log\det\left(\I_{N_k}+\frac{1}{\sigmatwo}\Gammak^a\Qk\right)-\tr{\Kkopt\Qk}.
\end{align}
Then, the optimal $\bQ_k$ can be obtained as \eqref{Optimal_Q1} referring to the proof of \cite[Theorem 3.6]{YLH15}, thus omitted here. This concludes the proof.

\bibliographystyle{IEEEtran}
\bibliography{Refabrv,EM_RISDMA_ref}

\end{document}